\documentclass[journal]{IEEEtran}

\usepackage{amssymb, amsthm, amsmath}
\usepackage{hyperref}
\usepackage{proof-at-the-end}
\usepackage{xparse}
\usepackage{algorithm}
\usepackage{times}
\usepackage{float}
\usepackage{fixltx2e}
\usepackage[pdftex]{graphicx}
\usepackage{graphicx}
\usepackage{subcaption}
\usepackage{bm}
\usepackage{color}
\usepackage{cite}
\usepackage{array}
\usepackage{url}
\usepackage{stackengine}
\newcommand\barbelow[1]{\stackunder[1.2pt]{$#1$}{\rule{.8ex}{.075ex}}}
\newtheorem{remark}{Remark}
\newtheorem{problem}{Problem}
\newtheorem{assumption}{Assumption}[section]
\newtheorem{thm}{Theorem}[section]
\newtheorem*{thm*}{Theorem}

\newtheorem*{lemma*}{Lemma}

\newtheorem*{proposition*}{Proposition}

\newtheorem{corollary}[thm]{Corollary}
\newtheorem*{corollary*}{Corollary}

\usepackage{color, soul}

\usepackage{xparse}
\NewDocumentEnvironment{mynormalthm}{O{}O{}+b}{%
  \begin{theoremEnd}[normal,#2]{thm}[#1]%
    #3%
  \end{theoremEnd}%
}{}

\pgfkeys{/prAtEnd/global custom defaults/.style={
one big link={See proof in Section~\ref{proofssection}.}
}
}
\newcommand{\update}[1]{{\color{black}#1}}

\usepackage{algorithm}
\usepackage{algpseudocode}
\newcommand*{\PICTURES}{}
\begin{document}
\title{\LARGE A Distributed Active Perception Strategy for Source Seeking and Level Curve Tracking }
\author{Said~Al-Abri,~\IEEEmembership{Member,~IEEE,}
        and~Fumin~Zhang,~\IEEEmembership{Senior Member,~IEEE}
\thanks{Said Al-Abri and Fumin Zhang are at School of Electrical and Computer
Engineering, Georgia Institute of Technology, Atlanta, GA 30332, E-mail:
{\tt\small \{saidalabri,fumin\}@gatech.edu}. 
}}
\IEEEoverridecommandlockouts
\IEEEpubid{\makebox[\columnwidth]{\copyright2020 IEEE \hfill} \hspace{\columnsep}\makebox[\columnwidth]{ }}

\maketitle
\IEEEpubidadjcol
\begin{abstract}                          
Algorithms for multi-agent systems to locate a source or to follow a desired level curve of spatially distributed scalar fields generally require sharing field measurements among the agents for gradient estimation. Yet, in this paper, we propose a distributed active perception strategy that enables swarms of various sizes and graph structures to perform source seeking and level curve tracking without the need to explicitly estimate the field gradient or explicitly share measurements.
The proposed method utilizes a consensus-like Principal Component Analysis perception algorithm that does not require explicit communication in order to compute a local body frame. 
This body frame is used to design a distributed control law where each agent modulates its motion based only on its instantaneous field measurement.  
Several stability results are obtained within a singular perturbation framework which justifies the convergence and robustness of the strategy. Additionally, efficiency is validated through various computer simulations and robots implementation in $2$-D scalar fields. The active perception strategy leverages the available local information and has the potential to be used in various applications such as modeling information propagation in biological and robotic swarms. 
\end{abstract}

\section{Introduction}
An important problem in swarm robotics is the deployment of multiple robots in order to achieve source seeking or level-curve tracking behavior in a scalar field. In source seeking problems, agents are tasked with finding the location that minimizes or maximizes the scalar field while in level-curve tracking, agents are tasked with tracking a trajectory that achieves a constant field value.  The field can represent environmental characteristics such as chemical concentrations, light intensities, or heat. These two problems have various applications including environmental monitoring, source signal localization, exploration, hazardous regions mapping, and search and rescue \cite{gupta2020development,lazna2018cooperation,mohan2009extensive,hutchinson2017review,khapalov2010source}.

In the literature, previous efforts to solve the dual problems of source seeking and level-curve tracking have relied on field gradient and Hessian estimation \cite{ogren2004cooperative,brinon2015distributed,fabbiano2016distributed,sakurama2016source,zhang2010cooperative,williams2012probabilistic,turgeman2017mission,barogh2017cooperative},   extremum seeking control \cite{cochran2009nonholonomic,cochran20093,khong2014multi,kvaternik2012analytic}, sliding-mode control \cite{ovchinnikov2014decentralized,matveev2015robot} , and weighted consensus laws \cite{elor2014robot}.  Most of the aforementioned control strategies rely either on sharing measurements via communication channels,  requiring specific spatial formations, or apply only to certain sizes and structures of interacting graphs. Communication-based solutions are particularly difficult to implement due to harsh environmental conditions that prevent networking such as in underwater mobile sensor networks \cite{kinsey2006survey}. 

In this paper, we propose a distributed strategy composed of two layers for perception and control.
In the perception layer, each agent uses the relative positions of its neighbors to learn a geometric body frame. In the control layer, each agent modulates its motion based on the body frame and a locally measured environmental field value. The interplay between the two layers results in an indirect active perception of the spatial gradient of the environmental property where the controlled  behavior of the agents enhances the information contents of the instantaneous measurements of the field and relative positions. This strategy enables swarms of various sizes and graph structures to perform collective source seeking and level curve tracking of scalar fields without the need to explicitly estimate the field gradient or explicitly exchange field measurements.

For source seeking,  in \cite{ogren2004cooperative}, the field gradient is assumed to be known and agents velocities are designed to climb the gradient and move in a desired formation.  In \cite{brinon2015distributed}, agents are required to form a circular formation, and then exchange field measurements via a communication channel to estimate and climb the gradient. Without knowing the global positions of the agents, an algorithm is developed in  \cite{fabbiano2016distributed}, however, it incorporates explicit sharing of field measurements to estimate the gradient. A different gradient-based strategy is presented in \cite{sakurama2016source} where agents autonomously split into multiple subgroups that each steer towards a source. 
Alternatively, extremum-based source seeking techniques are developed for a single vehicle in   $2$-D space in \cite{cochran2009nonholonomic}, and in  $3$-D space in \cite{cochran20093}. The proposed method relies on a constant forward velocity and designs the angular velocity in order to achieve extremum seeking behavior \cite{ariyur2003real}.  
Although the approach is simple to implement, the vehicle needs a relatively long distance to travel until the performance improves. A multi-agent extremum-based source seeking is developed in \cite{khong2014multi} and \cite{kvaternik2012analytic}, however, the agents need to exchange some estimated parameters. 
In \cite{elor2014robot},  a strategy is developed for a large number of robots with a complete graph based on a weighted consensus and a Gaussian perturbation. Although it is independent of communication,  agents bounce randomly in all directions leading to a slow  and impractical  movement towards the source.

For the level curve tracking, in \cite{zhang2010cooperative} and \cite{williams2012probabilistic}, the field gradient is assumed to be known. Alternatively, the algorithms in \cite{turgeman2017mission} and \cite{barogh2017cooperative} rely on communicating field measurements and maintaining prescribed formations to estimate the field gradient.
\update{In \cite{wang2019dynamic}, a cooperative control law is designed for two agents such that one agent estimates the field gradient and the other one tracks the plume front. }
Independent of gradient estimation,  \update{algorithms are designed in \cite{chatterjee2017cooperative,chatterjee2019modular} for a $2$-agent system but require} communicating field measurements. A discontinuous sliding mode control law is designed in \cite{ovchinnikov2014decentralized} and \cite{matveev2015robot} for an agent to track the level curve, independent of both gradient estimation and measurement communication, while a formation law is designed to spread the agents across the level curve.

Inspired by a school of fish seeking darker areas \cite{berdahl2013emergent}, the  Speeding-Up and Slowing-Down (SUSD) strategy is developed for source seeking without gradient estimation in  \cite{wu2015speeding} and \cite{wu2013bio} for 
$2$-D and $3$-D environments, respectively. The SUSD strategy requires a common motion direction that can be computed locally and without explicit communication only for $2$-agent and $3$-agent systems in $2$-D and $3$-D, respectively.  Differently, in   \cite{al2018gradient} we used  a leader-follower consensus-on-a sphere to obtain the common motion direction where agents are assumed to be able to measure the velocity directions of their neighbors.

The primary novelty of this work is in utilizing the Oja Principal Component Analysis (PCA)  flow \cite{oja1982simplified,baldi1995learning,yoshizawa2001convergence} to agree on a frame to describe the shape of the swarm, which we call a body frame. The PCA flow works as a consensus law, however, its input is the covariance of the relative positions. This is different than the common consensus law which requires the headings of neighbors as an input \cite{markdahl2017almost}. Since the relative positions are locally measured, then the PCA flow achieves consensus without requiring the agent to exchange data among them.

The main challenge this paper solves is to design a motion direction that all agents compute locally without communication. This motion direction  converges to the negative direction of the field gradient without estimating it. We solve this challenge by the PCA perception algorithm which, by capturing changes on the spatial shape and orientation of the swarm,  represents an indirect feedback signal of how the field is affecting the motions of other agents. Additionally, 
by autonomously tuning the velocity along the components of the body frame, the strategy achieves  both source seeking and level curve tracking with a single control law.

The first contribution of this paper is utilizing a PCA perception algorithm on relative positions to achieve a consensus in the body frame. The second contribution is a distributed control law that accomplishes both missions of source seeking and level curve tracking. The third contribution is deriving the information dynamics, not only for source seeking and level curve tracking but for general control laws. 
The fourth contribution is obtaining input-to-state stability results within a singular perturbation framework analysis 
for  (1) the convergence of the SUSD search direction to the negative gradient direction for source seeking and level curve tracking under complete graphs, (1) the convergence of the SUSD search direction to the negative gradient direction for source seeking under incomplete graphs
(3) the convergence of the swarm to the source location under complete and incomplete graphs, and to the desired level curve under complete graphs.
These results reflect robust convergence to both the source location under complete and incomplete graphs and to the desired level curve under complete graphs. The last contribution is validating the proposed strategy for various source seeking and level curve tracking behaviors through simulations and experiments. The experiments are conducted using the Georgia Teach Robotarium \cite{pickem2017robotarium} and the  Georgia Teach Miniature Autonomous Blimps \cite{cho2017autopilot}.

Preliminary results of this paper appear in our two conference papers \cite{al2018integrating,al2018distributed}. In these two papers, we only considered complete graphs, and in the convergence analysis, we ignored the higher-order terms of the field in both the dynamics derivation and convergence analysis. Significantly, in this paper, we derive the information dynamics for the general case of incomplete graphs and for a generic control law within the distributed active perception algorithm. 
Additionally, we consider the nonlinearities of the field which allows us to refine the convergence neighborhood around the desired equilibrium. While in the conference papers we only proved that the SUSD direction converges to the negative gradient direction, in this paper, we also prove that the swarm converges to either the source location or the desired level curve.  
Furthermore, the conference papers do not include any experimental results.  

The proposed distributed active perception strategy offers a new method that leverages the available local information and
enables robots with limited resources to perform various swarming activities. The PCA body frame might be used to design control laws for purposes other than the source seeking and level curve tracking. An important insight we analytically show in this paper is that the field measurements are communicated via the distributed perception algorithm. This might be useful in modeling information propagation in biological and robotic swarms \cite{mateo2017effect,haque2018analysis,mischiati2017geometric,daniels2016quantifying}.

The rest of the paper is organized as follows. The problem is formulated in Section~\ref{problem formulation}.   Then the
distributed active perception strategy is presented in  Section~\ref{Design}. In Section~\ref{Dynamics}, we derive the information dynamics which are used in  Section~\ref{Convergence} for stability analysis. Finally, simulation and experimental results are given in and Section~\ref{Results},  and concluding remarks and suggestions for future work are provided in Section~\ref{Conclusion}.

\section{Problem Formulation}\label{problem formulation}
\subsection{Preliminaries}
In this paper, we consider a swarm of $M$ agents described by an undirected visibility graph, $\mathcal{G}\subseteq\mathcal{V}\times \mathcal{E}$ where $\mathcal{V}$ is the set of all agents, and $\mathcal{E}$ is the set of all edges. An undirected edge $(i,j)\in\mathcal{E}$ exists if both agents can sense the relative positions of each other. A graph is connected if, for each $i,j\in\mathcal{V}$, there exists a sequence of edges connecting the $i$-th and $j$-th agents. If each agent shares an edge with all other agents, then the graph is \textbf{complete}, otherwise, it is \textbf{incomplete}. The neighboring set of $i$ is defined by $\mathcal{N}_i=\{j|(i,j)\in\mathcal{E}\}$. Additionally, if for each agent the neighboring set $\mathcal{N}_i$ is fixed,  the graph is static, otherwise, it is dynamic. We consider the following assumption about the graph.
\begin{assumption}\label{as:graph}
$\mathcal{G}$ is undirected and connected. 
\end{assumption}
This assumption is to simplify the convergence analysis. However, the design in Section~\ref{Design} is applicable to a broader class of graphs which will be supported by simulation results. Additionally, we will show in Section~\ref{Dynamics} that connectivity is implicitly guaranteed when the graph is complete.

Let  $\bm{r}_i\in \mathbb{R}^2$ be the position of the $i$-th agent in a $2$-D space. 
\begin{assumption}\label{as:relativepositions}
Each $i$-th agent knows the relative positions   $(\bm{r}_j-\bm{r}_i)$ of all its neighbors,  $j\in\mathcal{N}_i$.
\end{assumption}
In practice, robots can be equipped with sensors to measure the relative positions of their neighbors with respect to their local frame, which is less challenging than requiring the global positions \cite{oh2015survey}. 

Furthermore, suppose there exists a scalar field $z:\mathbb{R}^2\to\mathbb{R}$. The analytical expression of the field function $z$ is not known, but each agent can only measure its value  $z(\bm{r}_i)$ at its current position $\bm{r}_i(t)$.  
\begin{assumption}\label{fieldassumption}
The field is assumed to be \update{real analytic}, time-invariant and bounded, i.e. $0\leq z(\bm{r}_i)\leq z_{max}$, and has a unique minimum at the source location $\bm{r}_0$ where $z(\bm{r}_0)=0$. 
\end{assumption}
\update{The field being real analytic implies that it is smooth and at any location it can be approximated by a Taylor expansion. 
We require the smoothness of the field because later we are going to make the speed of each agent to be proportional to the field measurement. However, some non-smooth fields might be transformed into smooth fields using, for example, Stochastic modeling, as in \cite{wu2013bio}. Additionally, we require the use of Taylor approximation for the information dynamics derivations and the convergence analysis.}

Consider $z^d\in\mathbb{R}$ to be a desired level curve field value, where a level curve is the set $\{\bm{r}|z(\bm{r})=z^d,\forall r\in\mathbb{R}^2\}$.
\update{
\begin{assumption}
The desired level curve $\{\bm{r}|z(\bm{r})=z^d\}$ is  connected.
\end{assumption}
}

Finally, let the motion of each agent  be described by
\begin{align}\label{eq:velocity}
&\dot{\bm{r}}_i=\frac{d\bm{r}_i}{dt}=\bm{u}_i,\quad i=1,\cdots,M,
\end{align}
where $\bm{u}_i$ is a local control law to be designed. 
\subsection{Problem Statement}
\begin{problem}
Without explicitly estimating the field gradient  and without explicitly communicating field measurements, design the local control law  $\bm{u}_i$, such that the swarm  autonomously steers towards either the source location $\bm{r}_0$, or the desired level curve  $\{\bm{r}|z(\bm{r})=z^d,\forall r\in\mathbb{R}^2\}$, and keeps tracking it.
\end{problem}
\begin{problem}
Analyze the convergence and robustness of the proposed strategy.
\end{problem}
\begin{remark}
Note that in this paper we assume the measurements to be noiseless. However, early versions of SUSD, as in \cite{wu2015speeding},  show that it is robust against noisy measurements. We will consider analyzing the robustness of the proposed control laws in this paper against noisy measurements for future works.    
\end{remark}
\section{The Distributed Active Perception Strategy }\label{Design}
\begin{figure}[h]
\vspace{-15pt}
\centering
\includegraphics[scale=0.5]{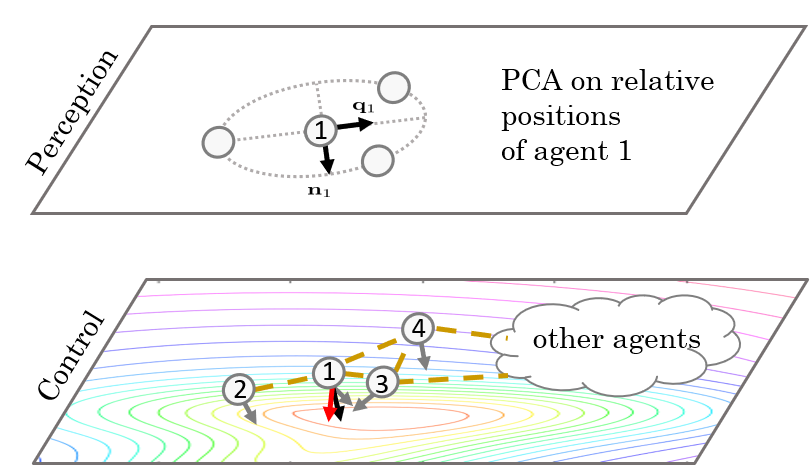}
\caption{The two layers of the active perception strategy. In the perception layer, agent $1$ learns the PCA body frame components $(\bm{q}_1,\bm{n}_1)$. Along these components agent $1$ modulate its motion based only on the instantaneous field measurement.   }\label{fig:multilayer}
\end{figure}

The proposed strategy is composed of two layers for perception and control as illustrated in Fig.~\ref{fig:multilayer}.
In the perception layer, each agent learns from the relative positions a time-varying locally computed body frame. The motion of each agent is designed in the control layer based on the perceived body frame and the locally measured environmental field value. The interplay between the two layers results in an active perception of the spatial gradient of the environmental property where the motions of the agents enhance the information contents of the instantaneous measurements of the field and relative positions.  In what follows, we first present the perception algorithm and then the distributed control law.
\vspace{-15pt}
\subsection{The PCA Perception Algorithm}
\begin{figure}[h]
\vspace{-10pt}
\centering
\includegraphics[scale=0.35]{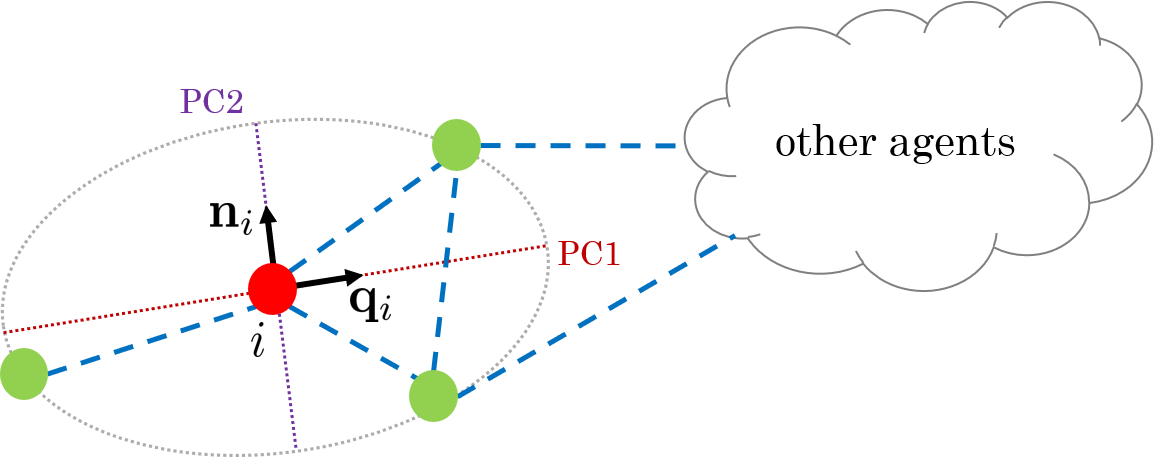}
\caption{The blue dash lines are the edges of the connectivity graph. The shape that agent $i$  observes is composed of itself and the $3$ neighbors from which agent $i$ learns the directions of the largest and smallest variances, PC1 and PC2. Across these variances, agent $i$ forms the body frame $(\bm{q}_i,\bm{n}_i)$.   }\label{fig:PCA_Comp}
\end{figure}

Principal Component Analysis (PCA) is a statistical method that computes directions of maximum (minimum) variation of a data set \cite{Jollifi02}. Given a covariance matrix of a data set, its eigenvector corresponding to the maximum (minimum) eigenvalue represents the direction of the maximum  (minimum) variance of the data, with the eigenvalue giving the variance of the data along that direction. For example, in Fig.~\ref{fig:PCA_Comp}, the data set of the positions of the agent $i$ and its neighbors (the spatial shape) has a maximum variance along the PC1 direction, and a minimum variance along the PC2 direction.

For each agent, consider the set of positions defined by $\mathcal{H}_i= \mathcal{N}_i\bigcup \{i\}$.
Then the position covariance matrix $\bm{C}_i(t)\in \mathbb{R}^{2\times 2}$ observed by each agent is constructed locally as
\begin{align}\label{eq:covariancematrix}
&\bm{C}_i(t)= \sum_{k\in \mathcal{H}_i }\big(\bm{r}_k(t) - \bm{r}_{c,i}(t)\big)\big(\bm{r}_k(t) - \bm{r}_{c,i}(t)\big)^\intercal,
\end{align}
where $\bm{r}_{c,i}=\frac{1}{M_i}\sum_{k\in \mathcal{H}_i }\bm{r}_k$ is the center of the swarm as seen by agent $i$, and $M_i=|\mathcal{H}_i|$. 

Let the PCA body-frame of agent $i$ be $(\bm{q}_i(t),\bm{n}_i(t))$, where $\bm{q}_i(t)$ and $\bm{n}_i(t)$ are orthonormal vectors in $ \mathbb{R}^2$  that represent the principal components of the covariance matrix $\bm{C}_i(t)$,  corresponding to the largest and smallest eigenvalues, $\lambda_i^q$ and $\lambda_i^n$, respectively.

Let $(\hat{\bm{q}}_i,\hat{\bm{n}}_i)$ be an estimate of the true PCA body-frame $(\bm{q}_i(t),\bm{n}_i(t))$, which is   given by the Oja PCA flow  \cite{baldi1995learning,yoshizawa2001convergence}
\begin{align}\label{eq:PCAflow}
&\frac{d\hat{\bm{q}}_i}{d\tau}=(\bm{I}-\hat{\bm{q}}_i\hat{\bm{q}}_i^\intercal)\bm{C}_i(t)\hat{\bm{q}}_i,\quad\hat{\bm{n}}_i=\bm{R}\hat{\bm{q}}_i,
\end{align}
where $\bm{R}$ is a $90^\circ$ counterclockwise rotation matrix.  Observe that we use the argument $\tau$ instead of $t$ to emphasize that for any given covariance matrix $\bm{C}_i(t)$ at time instant $t$, agent $i$ runs  (\ref{eq:PCAflow}) at a different time scale $\tau$. Since each agent may have different neighbors, as in the case when the graph is incomplete, then each agent may obtain different principal directions. Observe that the PCA  model  (\ref{eq:PCAflow}) is scalable to graphs with an arbitrary number of agents and structures. 

\begin{assumption}\label{shape assumption}
For each agent, $\lambda_i^q\neq\lambda_i^n$.
\end{assumption}
This is to ensure the eigenvectors $\bm{q}_i$ and $\bm{n}_i$ are orthogonal which is a requirement for mathematical correctness of the derived dynamics and convergence results obtained in this paper. Otherwise, such as when the agents are evenly distributed on a circle, the covariance matrix may have multiple solutions where some of them may not be orthogonal. Due to sensing errors on measuring relative positions, it is unlikely  to have $\lambda_i^q=\lambda_i^n$. Additionally, since we design $\hat{\bm{n}}_i=\bm{R}\hat{\bm{q}}_i$ in (\ref{eq:PCAflow}), then $\bm{q}_i$ and $\bm{n}_i$ are ensured to be orthogonal even if $\lambda_i^q=\lambda_i^n$. Moreover, at each time $t$, we initialize (\ref{eq:PCAflow}) with $\bm{q}_i(\tau_0)=\bm{q}_i(t-dt)$ so to ensure that the obtained solution is not arbitrary when $\lambda_i^q=\lambda_i^n$.

\begin{figure*}[tb]
\begin{subfigure}[b]{0.29\linewidth}
\includegraphics[scale=0.3]{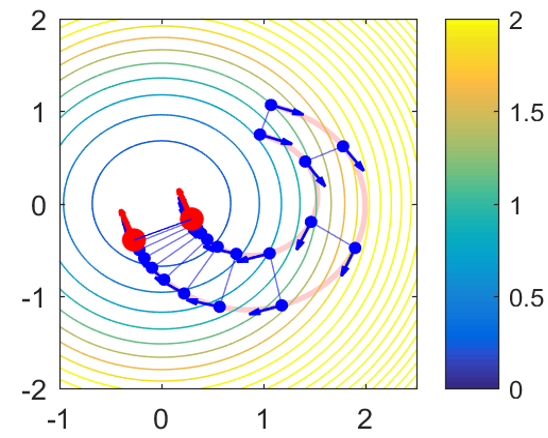}
\caption{$z^d=0$, $k_2=0$}
\end{subfigure}\qquad
\begin{subfigure}[b]{0.29\linewidth}
\includegraphics[scale=0.31]{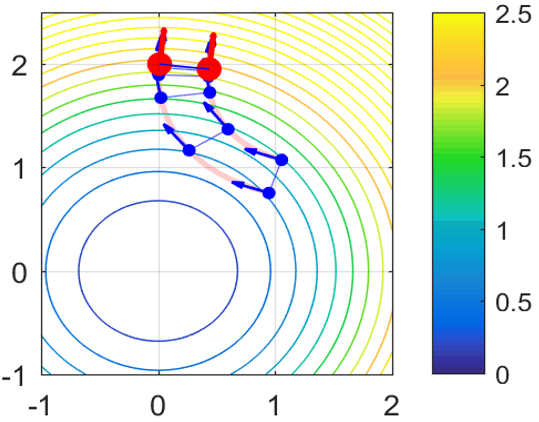}
\caption{$z^d\neq0$, $k_2=0$}
\end{subfigure}\qquad
\begin{subfigure}[b]{0.29\linewidth}
\includegraphics[scale=0.4]{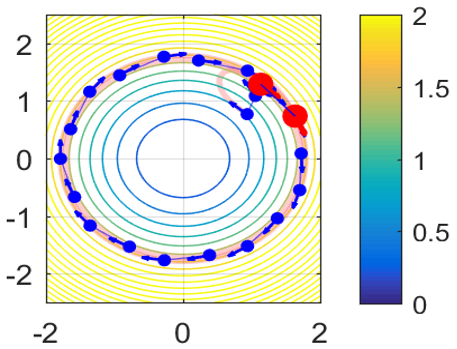}
\caption{$z^d\neq0$, $k_2\neq0$}
\end{subfigure}
\caption{The  blue arrows are the velocities which turn red at the end time. The circular curves are the field level curves.   }\label{fig:source_level_seeking}
\end{figure*}

\subsection{The Distributed Control Law}
Given the body frame $(\bm{q}_i(t),\bm{n}_i(t))$ obtained by  (\ref{eq:PCAflow}),  
we propose the control law
\begin{align}\label{eq:controllaw}
&\bm{u}_i(t)=k_1(z_i(t)-z^d)\bm{n}_i(t)+k_2\bm{q}_i(t),
\end{align}
where $z_i(t)$ and $z^d$ are measured and desired field values, respectively. The parameters $k_1,k_2\in \mathbb{R}$  are positive tuning parameters.
To intuitively explain  the control law (\ref{eq:controllaw}), we simulate it in Fig.~\ref{fig:source_level_seeking} for a $2$-agent system in a  scalar field. In this example, $\bm{q}_i$ is along the line-of-sight between the two agents, and $\bm{n}_i$  is perpendicular to the line-of-sight.  
When $z^d=0$ and $k_2=0$, then agent $i$ speeds up or slows down along the direction $\bm{n}_i$ depending on the local field measurement $z_i(t)$. Since the two agents are moving in the same direction at different speeds, then eventually they steer towards the minimum of the field. 

On the other hand, when $z^d\neq 0$ and $k_2= 0$, the two agents approach the level curve $\{\bm{r}|z(\bm{r}_1)=z(\bm{r}_2)=z^d\}$. Finally, when $z^d\neq 0$ and $k_2\neq 0$, then the first term $k_1(z_i(t)-z^d)\bm{n}_i(t)$ steers the 2-agent system towards the desired level curve, while the second term $k_2\bm{q}_i(t)$  moves the swarm along the level curve.  Note that the first term changes its sign as the sign of $(z_i(t)-z^d)$ changes, which  stabilize the agents at the level curve. Additionally, the gains $k_1$ and $k_2$ determine the tracking speed and  accuracy.  In particular, small $k_2$ compared to $k_1$ leads to slow tracking, but high accuracy, and vice versa. 

A pseudocode description for the  distributed active perception strategy  is given in Algorithm~\ref{alg:alg1}. The termination condition for source seeking could be to terminate when $z_i(t)<\bar{z}$ where $\bar{z}$ is a satisfactory threshold. For the level curve tracking, we may terminate after a certain desired time. 
All the steps in the algorithm are locally computed without explicit communication of any values. 
\begin{algorithm}
  \caption{The Distributed  Strategy}
  \begin{algorithmic}[1]
       \While{ termination condition is not met, }
     \For{each agent}
        \State compute covariance matrix using (\ref{eq:covariancematrix})\label{cov}
        \State compute principal components using (\ref{eq:PCAflow})\label{pc}
        \State update motion using (\ref{eq:controllaw})\label{pos}
      \EndFor
      \EndWhile
 \
  \end{algorithmic}
  \label{alg:alg1}
\end{algorithm}
\begin{remark}
The control law (\ref{eq:controllaw}) has three main differences than the source seeking laws in our previous works \cite{wu2013bio,wu2015speeding,al2018gradient}. First, it solves a combined task of source seeking and level curve tracking. Second, it does not require a rigid prescribed shape formation. Third, it utilizes a PCA perception algorithm to define a locally computed body frame, which previously is obtained by a consensus-on-a sphere law \cite{markdahl2017almost}.
\end{remark}
\begin{remark}
 Using a consensus law,  i.e. $\dot{\bm{q}}_i=(\bm{I}-\bm{q}_i\bm{q}_i^\intercal)\sum_{j\in\mathcal{N}_i}\bm{q}_j$  requires agents to share their $\bm{q}_i$ via  a communication channel. Interestingly,  the  PCA flow (\ref{eq:PCAflow}), i.e. $\frac{d\bm{q}_i}{d\tau}=(\bm{I}-\bm{q}_i(\tau)\bm{q}_i^\intercal(\tau))\bm{C}_i\bm{q}_i(\tau)$,  can also be viewed as a consensus law. However, its input is the local covariance $\bm{C}_i$, not the neighboring headings $\sum_{j\in\mathcal{N}_i}\bm{q}_j$.\update{ Therefore, agents can agree on a common direction by only observing local positions, without requiring them to communicate. As the agents change their speed due to the encountered field values, the PCA consensus value changes. Thus agents indirectly receive the effects of the field values on their neighbors by the change of the PCA value calculated locally. This cannot be achieved by using the classical heading average consensus where the consensus value is independent of the agents' positions.  
In fact, we will prove later that the PCA consensus value under some conditions aligns with the field gradient.}
\end{remark}

\begin{remark}
Extra terms can be added in  (\ref{eq:controllaw}) to, for example, maintain prescribed formations or avoid collisions. In the experimental results, we show examples of formation terms and their effects on the performance. However, we don't consider them in the convergence analysis for the sake of simplicity. Additionally, collision avoidance may be guaranteed using lower-level controllers such as barrier functions as in \cite{glotfelter2019hybrid}.  
\end{remark}

\section{ The Information Dynamics }\label{Dynamics}
Let $z_{c,i}^d=z_{c,i}-z^d$ where $z_{c,i}$ is the field measurement at the local center $\bm{r}_{c,i}=\frac{1}{M_i}\sum_{l\in\mathcal{H}_i}\bm{r}_{l}$. Define $\bm{N}_{c,i}=\nabla z(\bm{r}_{c,i})/\|\nabla z(\bm{r}_{c,i})\|$ to be a unit-length vector along the direction of the field  gradient at the local center $\bm{r}_{c,i}$. Then, using (\ref{eq:velocity}) and (\ref{eq:controllaw}), we obtain
\begin{align}\label{zcddot}
    &\dot{z}_{c,i}^d=\frac{\|\nabla z_{c,i}\|}{M_i}
    \sum_{l\in\mathcal{H}_i}[k_1(z_l-z^d)\langle\bm{N}_{c,i},\bm{n}_l\rangle+k_2\langle\bm{N}_{c,i},\bm{q}_l\rangle].
\end{align}
Observe that   $z_{c,i}^d\to0$ if and only if $z_{c,i}\to z^d$ for the level curve tracking, or $z_{c,i}\to 0$ for the source seeking. However, to analyze the convergence of the origin $z_{c,i}^d=0$ of (\ref{zcddot}), we need the dynamics of the  two principle directions  $(\dot{\bm{q}}_l,\dot{\bm{n}}_l)=(\frac{d\bm{q}_l}{dt},\frac{d\bm{n}_l}{dt})$ for $l=1,\cdots,M$.
\begin{remark}
The dynamics  $(\frac{d\bm{q}_l}{dt},\frac{d\bm{n}_l}{dt})$ for $l=1,\cdots,M$ are different from  $(\frac{d\hat{\bm{n}}_l}{d\tau},\frac{d\hat{\bm{q}}_l}{d\tau})$ given by
the  PCA flow  (\ref{eq:PCAflow}), which describes the dynamics of learning the body frame from   a given covariance matrix $\bm{C}_l(t)$ at time instant $t$.
\end{remark}

In what follows we  derive the dynamics of the body frame $(\dot{\bm{q}}_l,\dot{\bm{n}}_l)$ first for a general control law $\bm{u}_l$, and then for the proposed control law (\ref{eq:controllaw}). We start the derivation with  incomplete graphs and then we consider  complete graphs.  
These dynamics eventually will be used in Section~\ref{Convergence} for the convergence analysis of the proposed strategy.

Let the covariance matrix seen by agent $i$ be given by
(\ref{eq:covariancematrix}) with  $\lambda_i^n$ and $\lambda_i^q$ be the smallest and largest eigenvalues of $\bm{C}_i$, corresponding to the eigenvectors $\bm{n}_i$ and $\bm{q}_i$, respectively.
The following result presents the 
dynamics of the PCA body frame for a general control law, $\bm{u}_i$.

\begin{theoremEnd}{lemma}\label{LemmaGeneralControl} \update{Let Assumption~\ref{shape assumption} holds.}
Then, under incomplete graphs, and when the agents move according to (\ref{eq:velocity}),   the  dynamics of the body frame are
\vspace{-10pt}
\begin{align}\label{eq:nqdotgneralcontrol}
&\dot{\bm{n}}_i=-\kappa\bm{q}_i,\quad\dot{\bm{q}}_i=\kappa\bm{n}_i,\vspace{-10pt}
\end{align}
where \update{$\kappa=\frac{1}{\lambda_i^q-\lambda_i^n}
\sum_{k\in\mathcal{H}_i}
\langle\bm{q}_i ,\bm{u}_k\rangle\langle\bm{r}_k-\bm{r}_{c,i},\bm{n}_i\rangle
+\frac{1}{\lambda_i^q-\lambda_i^n}
\sum_{k\in\mathcal{H}_i}\langle\bm{q}_i,\bm{r}_k-\bm{r}_{c,i}\rangle\langle\bm{u}_k,\bm{n}_i\rangle$.}
\end{theoremEnd}
\begin{proofEnd}

The covariance matrix satisfies $\bm{C}_i\bm{n}_i=\lambda_i^n\bm{n}_i$, and $\bm{C}_i\bm{q}_i=\lambda_i^q\bm{q}_i$. \update{Since with Assumption~\ref{shape assumption}, the eigenvalues and eigenvectors of $\bm{C}_i$ are uniquely defined, then their derivatives exist.}
Thus, taking the derivative, we obtain
\begin{align}\label{eq:cndot22}
&\dot{\bm{C}}_i\bm{n}_i+\bm{C}_i\dot{\bm{n}}_i=\dot{\lambda}_i^n\bm{n}_i+\lambda_i^n\dot{\bm{n}}_i,
\end{align}
\begin{align}\label{eq:qndot22}
&\dot{\bm{C}}_i\bm{q}_i+\bm{C}_i\dot{\bm{q}}_i=\dot{\lambda}_i^q\bm{q}_i+\lambda_i^q\dot{\bm{q}}_i.
\end{align}
Inner product both sides of (\ref{eq:cndot22}) with the eigenvector $\bm{q}_i$, and both sides of (\ref{eq:qndot22}) with the eigenvector $\bm{n}_i$
\begin{align}\label{eqxxwen}
&\langle\bm{q}_i, \dot{\bm{C}}_i\bm{n}_i\rangle+\langle\bm{q}_i,\bm{C}_i\dot{\bm{n}}_i\rangle
=\dot{\lambda}_i^n\langle\bm{q}_i,\bm{n}_i\rangle+\lambda_i^n\langle\bm{q}_i,\dot{\bm{n}}_i\rangle,
\end{align}
\begin{align}\label{eqxxweq}
&\langle\bm{n}_i, \dot{\bm{C}}_i\bm{q}_i\rangle+\langle\bm{n}_i,\bm{C}_i\dot{\bm{q}}_i\rangle
=\dot{\lambda}_i^q\langle\bm{n}_i,\bm{q}_i\rangle+\lambda_i^q\langle\bm{n}_i,\dot{\bm{q}}_i\rangle.
\end{align}
Since $\bm{C}_i$ is symmetric, then $\dot{\bm{C}}_i$ is also symmetric. This implies that $\langle\bm{q}_i,\bm{C}_i\dot{\bm{n}}_i\rangle=\langle\bm{C}_i\bm{q}_i,\dot{\bm{n}}_i\rangle=\lambda_i^q\langle\bm{q}_i,\dot{\bm{n}}_i\rangle$. Similarly, $\langle\bm{n}_i,\bm{C}_i\dot{\bm{q}}_i\rangle=\langle\bm{C}_i\bm{n}_i,\dot{\bm{q}}_i\rangle=\lambda_i^n\langle\bm{n}_i,\dot{\bm{q}}_i\rangle$. 
Using these along with the fact that $\langle\bm{q}_i,\bm{n}_i\rangle=\langle\bm{n}_i,\bm{q}_i\rangle=0$,  we obtain from (\ref{eqxxwen}) and  (\ref{eqxxweq})
\begin{align}
&\langle\bm{q}_i,\dot{\bm{n}}_i\rangle=-\frac{1}{\lambda_i^q-\lambda_i^n}\langle\bm{q}_i, \dot{\bm{C}}_i\bm{n}_i\rangle,\label{eqxxn}\\
&\langle\bm{n}_i,\dot{\bm{q}}_i\rangle=\frac{1}{\lambda_i^q-\lambda_i^n}\langle\bm{q}_i, \dot{\bm{C}}_i\bm{n}_i\rangle.\label{eqxxq}
\end{align}
Since $\bm{n}_i$ and $\bm{q}_i$ are orthonormal, then we can write
\begin{align}\label{eq:nqdot5}
&\dot{\bm{n}}_i=\langle\bm{q}_i  ,\dot{\bm{n}}_i \rangle\bm{q}_i,\quad
\dot{\bm{q}}_i=\langle\bm{n}_i  ,\dot{\bm{q}}_i  \rangle\bm{n}_i.
\end{align}
Substituting (\ref{eqxxn}) and  (\ref{eqxxq}) in  (\ref{eq:nqdot5})
\begin{align}\label{eq:ndot5}
&\dot{\bm{n}}_i=-\frac{1}{\lambda_i^q-\lambda_i^n}\langle\bm{q}_i, \dot{\bm{C}}_i\bm{n}_i\rangle\bm{q}_i
\end{align}
\begin{align}\label{eq:qdot5}
&\dot{\bm{q}}_i=\frac{1}{\lambda_i^q-\lambda_i^n}\langle\bm{q}_i, \dot{\bm{C}}_i\bm{n}_i\rangle\bm{n}_i.
\end{align}
Taking the derivative  of the covariance (\ref{eq:covariancematrix}), we obtain
\begin{align}\label{ste32}
    &\dot{\bm{C}}_i=\sum_{k\in\mathcal{H}_i}\notag\\
    &
    [(\dot{\bm{r}}_k-\dot{\bm{r}}_{c,i})(\bm{r}_k-\bm{r}_{c,i})^\intercal+(\bm{r}_k-\bm{r}_{c,i})(\dot{\bm{r}}_k-\dot{\bm{r}}_{c,i})^\intercal].
\end{align}
But using (\ref{eq:velocity})
\update{
\begin{align}\label{an42}
    &\dot{\bm{r}}_k-\dot{\bm{r}}_{c,i}=\bm{u}_k-\frac{1}{M_i}\sum_{l\in{\mathcal{H}_i}}\bm{u}_l=
    \frac{1}{M_i}\sum_{l\in{\mathcal{H}_i}}(\bm{u}_k-\bm{u}_l).
\end{align}
}
Hence, substituting (\ref{an42}) in (\ref{ste32})
\update{
\begin{align}\label{covariance derivative}
    &\dot{\bm{C}}_i=\frac{1}{M_i}\sum_{k\in\mathcal{H}_i}
    \sum_{l\in{\mathcal{H}_i}}(\bm{u}_k-\bm{u}_l)(\bm{r}_k-\bm{r}_{c,i})^\intercal\notag\\
    &+\frac{1}{M_i}\sum_{k\in\mathcal{H}_i}
    (\bm{r}_k-\bm{r}_{c,i})\sum_{l\in{\mathcal{H}_i}}(\bm{u}_k-\bm{u}_l)^\intercal\notag\\
    &=\sum_{k\in\mathcal{H}_i}\big[\bm{u}_k(\bm{r}_k-\bm{r}_{c,i})^\intercal+\big(\bm{r}_k-\bm{r}_{c,i})\bm{u}_k^\intercal],
\end{align}
where we used the fact that
$\frac{1}{M_i}\sum_{k\in\mathcal{H}_i}
    \sum_{l\in{\mathcal{H}_i}}\bm{u}_k(\bm{r}_k-\bm{r}_{c,i})^\intercal=\sum_{k\in\mathcal{H}_i}\bm{u}_k(\bm{r}_k-\bm{r}_{c,i})^\intercal$,
$\frac{1}{M_i}\sum_{k\in\mathcal{H}_i}
    \sum_{l\in{\mathcal{H}_i}}\bm{u}_l(\bm{r}_k-\bm{r}_{c,i})^\intercal=
    (\frac{1}{M_i}\sum_{l\in{\mathcal{H}_i}})(\sum_{k\in\mathcal{H}_i}(\bm{r}_k-\bm{r}_{c,i})^\intercal)=0$,
    $\frac{1}{M_i}\sum_{k\in\mathcal{H}_i}
    (\bm{r}_k-\bm{r}_{c,i})\sum_{l\in{\mathcal{H}_i}}\bm{u}_k^\intercal=
    \sum_{k\in\mathcal{H}_i}
    (\bm{r}_k-\bm{r}_{c,i})\bm{u}_k^\intercal$,
    and $\frac{1}{M_i}\sum_{k\in\mathcal{H}_i}
    (\bm{r}_k-\bm{r}_{c,i})\sum_{l\in{\mathcal{H}_i}}\bm{u}_l^\intercal=0$.
}
Substituting (\ref{covariance derivative}) in (\ref{eq:ndot5}) and (\ref{eq:qdot5}) leads to the desired result (\ref{eq:nqdotgneralcontrol}).

\end{proofEnd}
It is interesting to observe that the actions of the neighbors \update{ in $\bm{u}_k$} are present in (\ref{eq:nqdotgneralcontrol}) not due to communication, but due to the distributed active perception algorithm where the body frame is obtained via the PCA  (\ref{eq:PCAflow}).  

We then derive the dynamics of the body frame under the proposed control law (\ref{eq:controllaw}).
\begin{theoremEnd}{lemma}
\update{Let Assumption~\ref{shape assumption} holds.} Then, using the motion dynamics (\ref{eq:velocity}) along with the control law (\ref{eq:controllaw}), the  dynamics of the body frame for source seeking and level curve tracking  with  general incomplete graphs are
\begin{align}
&\dot{\bm{n}}_i=-k_1\frac{1}{\lambda_i^q-\lambda_i^n}\bm{w}_i^\intercal\bm{q}_i\bm{q}_i-\frac{1}{\lambda_i^q-\lambda_i^n}\bm{\sigma}_i\bm{q}_i,\label{eq:ndot6}\\
&\dot{\bm{q}}_i=k_1\frac{1}{\lambda_i^q-\lambda_i^n}\bm{w}_i^\intercal\bm{q}_i\bm{n}_i+\frac{1}{\lambda_i^q-\lambda_i^n}\bm{\sigma}_i\bm{n}_i,\label{eq:qdot6}
\end{align}
where \update{for $z_k^d=z_k-z^d$ and $z_{c,i}^d=z_{c,i}-z^d$, }
\begin{align}
    &\bm{w}_i=\sum_{k\in\mathcal{H}_i}(z_k^d\langle \bm{n}_k,\bm{n}_i\rangle-z_{c,i}^d)(\bm{r}_k-\bm{r}_{c,i}),\label{wi}\\
    &\bm{\sigma}_i=k_1\sum_{k\in\mathcal{H}_i}(z_k^d\langle \bm{n}_k,\bm{q}_i\rangle-z_{c,i})\langle\bm{r}_k-\bm{r}_{c,i},\bm{n}_i\rangle+\notag\\
    &k_2\sum_{k\in\mathcal{H}_i}[\langle\bm{q}_k,\bm{n}_i\rangle \langle\bm{r}_k-\bm{r}_{c,i},\bm{q}_i\rangle+
    \langle\bm{q}_k,\bm{q}_i\rangle \langle\bm{r}_k-\bm{r}_{c,i},\bm{n}_i\rangle].\label{sigmaioriginal}
\end{align}
\end{theoremEnd}
\begin{proofEnd}
Substituting the control law (\ref{eq:controllaw}) in (\ref{covariance derivative}),  we obtain
\update{
\begin{align}
    &\dot{\bm{C}}_i=\sum_{k\in\mathcal{H}_i}\Big[(k_1z_k^d\bm{n}_k+k_2\bm{q}_k)(\bm{r}_k-\bm{r}_{c,i})^\intercal\notag\\
    &
    +(\bm{r}_k-\bm{r}_{c,i})(k_1z_k^d\bm{n}_k+k_2\bm{q}_k)^\intercal\Big],\label{'Cidotnewfinal'}
\end{align}
where $z_k^d=z_k-z^d$. Let $z_{c,i}^d=z_{c,i}-z^d$. 
Then, we add $0=-k_1\sum_{k\in\mathcal{H}_i}z_{c,i}^d\bm{n}_i(\bm{r}_k-\bm{r}_{c,i})^\intercal=k_1\sum_{k\in\mathcal{H}_i}z_{c,i}^d(\bm{r}_k-\bm{r}_{c,i})\bm{q}_i^\intercal$ to (\ref{'Cidotnewfinal'}) to obtain
\begin{align}
    &\dot{\bm{C}}_i=\sum_{k\in\mathcal{H}_i}\Big[\big(k_1[z_k^d\bm{n}_k-z_{c,i}^d\bm{n}_i]+k_2\bm{q}_k\big)(\bm{r}_k-\bm{r}_{c,i})^\intercal\notag\\
    &
    +(\bm{r}_k-\bm{r}_{c,i})\big(k_1[z_k^d\bm{n}_k-z_{c,i}^d\bm{q}_i]+k_2\bm{q}_k\big)^\intercal\Big],\label{'Cidotnewfinal2'}
\end{align}
which is used to derive
}
\begin{align}\label{nicidotqi2}
    &\langle \bm{n}_i,\dot{\bm{C}}_i\bm{q}_i\rangle=\sum_{k\in\mathcal{H}_i}\notag\\
    &[k_1(z_k^d\langle \bm{n}_k,\bm{n}_i\rangle-z_{c,i}^d)+k_2\langle\bm{q}_k,\bm{n}_i\rangle]\langle\bm{r}_k-\bm{r}_{c,i},\bm{q}_i\rangle+\notag\\
    &[k_1(z_k^d\langle \bm{n}_k,\bm{q}_i\rangle-z_{c,i}^d)+k_2\langle\bm{q}_k,\bm{q}_i\rangle]\langle\bm{r}_k-\bm{r}_{c,i},\bm{n}_i\rangle.
\end{align}
Finally, we obtain (\ref{eq:ndot6}) and (\ref{eq:qdot6}) by first substituting (\ref{nicidotqi2}) in (\ref{eq:ndot5}) and (\ref{eq:qdot5}), and then applying  (\ref{wi}) and (\ref{sigmaioriginal}) for $\bm{w}_i$ and $\bm{\sigma}_i$, respectively.

\end{proofEnd}
Note that $\bm{\sigma}_i\to\bm{0}$ as $\langle \bm{n}_i,\bm{n}_k \rangle\to1$ $\forall k\in\mathcal{H}_i$, i.e. as when the local PCA vectors align. 

To find the relationship between $\bm{n}_i$ and the local gradient $\nabla z_{c,i}$, we approximate the measurement $z_k=z(\bm{r}_k)$ by Taylor expansion with respect to the center $\bm{r}_{c,i}$. \update{Since according to  Assumption~\ref{fieldassumption} the field function $z$ is analytic, then we can write   }
\begin{align}\label{eq:field2}
    &z_k-z_{c,i}=\langle \bm{r}_k-\bm{r}_{c,i},\nabla z_{c,i}\rangle+\nu_k,
\end{align}
where $\nabla z_{c,i}=\nabla z(\bm{r}_{c,i})$ is the local gradient in the vicinity of the center $\bm{r}_{c,i}$ observed by agent $i$, and  $\nu_k=\mathcal{O}\|\bm{r}_k-\bm{r}_{c,i}\|$ represents the higher-order terms. \update{Note that for the field to be real analytic as required in Assumption~\ref{fieldassumption}, the distance $\|\bm{r}_k-\bm{r}_{c,i}\|$ has to be small enough such that  $z(\bm{r}_k)=\sum_{a=0}^\infty\frac{z^{(a)}(\bm{r}_{c,i})}{a!}(\bm{r}_k-\bm{r}_{c,i})^a$, where $z^{(a)}(\bm{r}_{c,i})$ is the $a-$th order derivative of $z$ at location $(\bm{r}_{c,i})$. Otherwise, $\nu_k$ in (\ref{eq:field2}) contains a residual term in addition to the higher-order derivatives of the field function. }

\begin{theoremEnd}{corollary}\label{corolaryincomplete}
\update{Let Assumption~\ref{fieldassumption} and Assumption~\ref{shape assumption} hold.} Then, using the motion dynamics (\ref{eq:velocity}) along with the control law (\ref{eq:controllaw}), the  dynamics of the body frame for source seeking with incomplete graphs are
\begin{align}
&\dot{\bm{n}}_i=-k_1\|\nabla z_{c,i}\|\frac{\lambda_i^q}{\lambda_i^q-\lambda_i^n}\langle\bm{N}_{c,i},\bm{q}_i \rangle\bm{q}_i-\hat{\nu}_i\bm{q}_i-\mathcal{E}_i\bm{q}_i,\label{eq:ndot888}\\
&\dot{\bm{q}}_i=+k_1\|\nabla z_{c,i}\|\frac{\lambda_i^q}{\lambda_i^q-\lambda_i^n}\langle\bm{N}_{c,i},\bm{q}_i\rangle \bm{n}_i+ \hat{\nu}_i\bm{n}_i
+\mathcal{E}_i\bm{n}_i,\label{eq:qdot888}
\end{align}
where $\bm{N}_{c,i}=\nabla z_{c,i}/\|\nabla z_{c,i}\|$,
$\hat{\nu}_i=(k_1/(\lambda_i^q-\lambda_i^n)\sum_{k\in\mathcal{H}_i}\nu_k \langle\bm{r}_k-\bm{r}_{c,i},\bm{q}_i\rangle$ is due to the nonlinearity of the field, 
and $\mathcal{E}_i=(k_1/(\lambda_i^q-\lambda_i^n))\sum_{k\in\mathcal{H}_i}z_k[
\langle \bm{n}_k,\bm{q}_i\rangle\langle\bm{r}_k-\bm{r}_{c,i},\bm{n}_i\rangle+(\langle \bm{n}_k,\bm{n}_i\rangle-1) \langle\bm{r}_k-\bm{r}_{c,i},\bm{q}_i\rangle]$ is due to the mismatch on the local PCA components.
\end{theoremEnd}
\begin{proofEnd}
Since we this lemma is about source seeking only, then we have $k_2=0$ and $z_{c,i}^d=z_{c,i}$.  
Substituting $z_{c,i}$ from  (\ref{eq:field2}) into (\ref{wi}), we obtain 
\begin{align}
    &\bm{w}_i=\bm{C}_i\nabla z_{c,i}+\sum_{k\in\mathcal{H}_i}[z_k(\langle \bm{n}_k,\bm{n}_i\rangle-1)+\nu_k](\bm{r}_k-\bm{r}_{c,i}),
\end{align}
where $\bm{C}_i\nabla z_{c,i}=\sum_{k\in\mathcal{H}_i}(\bm{r}_k-\bm{r}_{c,i})(\bm{r}_k-\bm{r}_{c,i})^\intercal\nabla z_{c,i}$.
Then we obtain
\begin{align}\label{ggh1}
    &k_1\frac{1}{\lambda_i^q-\lambda_i^n}\langle\bm{w}_i,\bm{q}_i\rangle=
    k_1\frac{\lambda_i^q}{\lambda_i^q-\lambda_i^n}\|\nabla z_{c,i}\|\langle\bm{N}_{c,i},\bm{q}_i\rangle+\hat{\nu_i} \notag\\
    &-k_1\frac{\lambda_i^q}{\lambda_i^q-\lambda_i^n}\sum_{k\in\mathcal{H}_i}z_k(\langle \bm{n}_k,\bm{n}_i\rangle-1)\langle\bm{r}_k-\bm{r}_{c,i}\bm{q}_i\rangle,
\end{align}
where $\hat{\nu}_i=(k_1/(\lambda_i^q-\lambda_i^n)\sum_{k\in\mathcal{H}_i}\nu_k \langle\bm{r}_k-\bm{r}_{c,i},\bm{q}_i\rangle$.
On the other hand, when $k_2=0$, then using (\ref{sigmaioriginal}), we obtain
\begin{align}\label{ggh2}
    &\frac{1}{\lambda_i^q-\lambda_i^n}\bm{\sigma}_i=\frac{k_1}{\lambda_i^q-\lambda_i^n}\sum_{k\in\mathcal{H}_i}z_k\langle \bm{n}_k,\bm{q}_i\rangle\langle\bm{r}_k-\bm{r}_{c,i},\bm{n}_i\rangle,
\end{align}
where we use the fact that $\sum_{k\in\mathcal{H}_i}z_{c,i}\langle\bm{r}_k-\bm{r}_{c,i},\bm{n}_i\rangle=z_{c,i}\bm{n}_i^\intercal\sum_{k\in\mathcal{H}_i}(\bm{r}_k-\bm{r}_{c,i})=0$.
Finally, substituting (\ref{ggh1}) and (\ref{ggh2}) into 
(\ref{eq:ndot6}) and (\ref{eq:qdot6}), and defining 
$\mathcal{E}_i=(k_1/(\lambda_i^q-\lambda_i^n))\sum_{k\in\mathcal{H}_i}z_k[
\langle \bm{n}_k,\bm{q}_i\rangle\langle\bm{r}_k-\bm{r}_{c,i},\bm{n}_i\rangle+(\langle \bm{n}_k,\bm{n}_i\rangle-1) \langle\bm{r}_k-\bm{r}_{c,i},\bm{q}_i\rangle]$,
we obtain the claimed (\ref{eq:ndot888}) and (\ref{eq:qdot888}).
\end{proofEnd}
Observe that this result is only for source seeking. For the level curve tracking, a similar result can be obtained with a different $\mathcal{E}_i$. We omit this case as in this paper we don't analyze the convergence of level curve tracking when the graph is incomplete.

When the graph is complete, we  view the entire swarm as a super agent and define $\nabla z_c=\nabla z(\bm{r}_c)$ to be the field gradient at the center of the swarm. Then, we obtain the following result. 
\begin{theoremEnd}{lemma}\label{lemmabodyframeboth}
\update{Let Assumption~\ref{fieldassumption} and Assumption~\ref{shape assumption} hold.}
Then, using the motion dynamics (\ref{eq:velocity}) along with the  control law (\ref{eq:controllaw}), the  dynamics of the body frame for source seeking and level curve tracking with  complete graphs are
\begin{align}
&\dot{\bm{n}}
=-k_1\|\nabla z_c\|\frac{\lambda^q}{\lambda^q-\lambda^n}\langle \bm{N}_c,\bm{q} \rangle\bm{q}
-\hat{\nu}\bm{q},\label{eq:ndot6completelinearizedx}\\
&\dot{\bm{q}}
=+k_1\|\nabla z_c\|\frac{\lambda^q}{\lambda^q-\lambda^n}\langle \bm{N}_c,\bm{q} \rangle\bm{n}
+\hat{\nu}\bm{n},\label{eq:qdot6completelinearizedx}
\end{align} 
where $\bm{N}_c=\frac{\nabla z_c}{\|\nabla z_c\|}$, and $\hat{\nu}=\frac{k_1}{\lambda^q-\lambda^n}\sum_{k=1}^M\nu_k\langle\bm{r}_k-\bm{r}_c,\bm{q}\rangle$.
\end{theoremEnd}
\begin{proofEnd}
We first proof that
\begin{align}
&\dot{\bm{n}}=-\frac{k_1}{\lambda^q-\lambda^n}\bm{w}^\intercal\bm{q}\bm{q}\label{eq:ndot6complete}\\
&\dot{\bm{q}}=+\frac{k_1}{\lambda^q-\lambda^n}\bm{w}^\intercal\bm{q}\bm{n}\label{eq:qdot6complete},
\end{align}
where
\begin{align}\label{wcomplete}
    &\bm{w}=\sum_{k=1}^M(z_k-z_c)(\bm{r}_k-\bm{r}_c).
\end{align}
When the graph is complete, then each agent computes the same covariance matrix
\begin{align}\label{covariancematrixcomplete}
&\bm{C}_i=\bm{C}=\sum_{k=1}^M(\bm{r}_k-\bm{r}_c)(\bm{r}_k-\bm{r}_c)^\intercal,
\end{align}
where all the agents see the same  center $\bm{r}_c=\frac{1}{M}\sum_{k=1}^M\bm{r}_k$.
 This implies that $\bm{n}_i=\bm{n}_k=\bm{n}$, and $\bm{q}_i=\bm{q}_k=\bm{n}$  for all $i,k$. 
 Hence, $\langle\bm{n}_i,\bm{q}_k\rangle=0$, and $\langle\bm{q}_i,\bm{q}_k\rangle=1$ for all $i,k$, and thus
  \begin{align}\label{sigmacomplete}
    &\bm{\sigma}_i=\sum_{k\in\mathcal{H}_i}[-k_1z_{c,i}^d\langle\bm{r}_k-\bm{r}_{c,i},\bm{n}_i\rangle+k_2 \langle\bm{r}_k-\bm{r}_{c,i},\bm{n}_i\rangle]=0,
\end{align}
 where we used the fact that $\sum_{k\in\mathcal{H}_i}(\bm{r}_k-\bm{r}_{c,i})=0$. Additionally,
 since $z_k^d-z_c^d=z_k-z^d-\frac{1}{M}\sum_{k=1}^M(z_k-z^d)=z_k-z_c$, and $\langle\bm{n}_k ,\bm{n}_i \rangle=1$, then (\ref{wi}) we obtain 
 (\ref{wcomplete}).
 Finally, using (\ref{sigmacomplete}), and substituting (\ref{wcomplete})  in (\ref{eq:ndot6}) and (\ref{eq:qdot6}) yields the claimed
(\ref{eq:ndot6complete}) and (\ref{eq:qdot6complete}).

Substituting $(z_k-z_c)$ of (\ref{eq:field2}) in (\ref{wcomplete}) yields
\begin{align}\label{www}
    &\bm{w}=\sum_{k=1}^M[\langle \bm{r}_k-\bm{r}_c,\nabla z_c\rangle(\bm{r}_k-\bm{r}_c)+\nu_k(\bm{r}_k-\bm{r}_c)].
\end{align}
But, $\sum_{k=1}^M\langle \bm{r}_k-\bm{r}_c,\bm{\nabla z_c}\rangle(\bm{r}_k-\bm{r}_c)=\sum_{k=1}^M(\bm{r}_k-\bm{r}_c)(\bm{r}_k-\bm{r}_c)^\intercal\nabla z_c=\bm{C}\nabla z_c$. Hence
\begin{align}\label{wicomplete}
    &\bm{w}=\bm{C}\nabla z_c+\sum_{k=1}^M\nu_k(\bm{r}_k-\bm{r}_c).
\end{align}
Finally, using $\hat{\nu}=\frac{k_1}{\lambda^q-\lambda^n}\sum_{k=1}^M\nu_k\langle\bm{r}_k-\bm{r}_c,\bm{q}\rangle$, and substituting (\ref{wicomplete}) in (\ref{eq:ndot6complete}) and (\ref{eq:qdot6complete}) yields the claimed (\ref{eq:ndot6completelinearizedx}) and  (\ref{eq:qdot6completelinearizedx}).
\end{proofEnd}
Note that, although when the graph is complete, it is not enough to  substitute $\mathcal{E}_i=0$ in  \textbf{Corollary~\ref{corolaryincomplete}} to obtain the result in \textbf{Lemma~\ref{lemmabodyframeboth}}, as \textbf{Corollary~\ref{corolaryincomplete}} is valid only for source seeking. 

Since $\bm{n}$ and $\bm{q}$ are orthonormal, we can write $\bm{q}\bm{q}^\intercal=\bm{I}-\bm{n}\bm{n}^\intercal$. Hence, we can reform  (\ref{eq:ndot6completelinearizedx}) as
\begin{align}\label{eq:ndot6complete22}
&\dot{\bm{n}}=-k_1\frac{\lambda^q}{\lambda^q-\lambda^n}\|\nabla z_c\|(\bm{I}-\bm{n}\bm{n}^\intercal)\bm{N}_c-\hat{\nu}\bm{q}.
\end{align} 
Note that the second term in (\ref{eq:ndot6complete22}) vanishes when $\nu_k=\nu$ for all agents. i.e. the field is linear, or when the agents are at the same level curve.
\begin{remark}
The first term   in (\ref{eq:ndot6complete22}) represents a consensus-on-a sphere control law \cite{markdahl2017almost}. This is interesting since although we are explicitly
applying (\ref{eq:controllaw}) with (\ref{eq:PCAflow}), the direction $\bm{n}$ is implicitly tracking the negative direction of the gradient $-\bm{N}_c$. 
\end{remark}

We conclude this section by the following result
\begin{theoremEnd}{lemma}\label{'lambdaqdot is zero'} 
\update{Let Assumption~\ref{fieldassumption} and Assumption~\ref{shape assumption} hold.}
Suppose the graph is complete and let  the motion dynamics of each agent be described by (\ref{eq:velocity}). Then, for both source seeking and level curve tracking,  $\lambda^n(t)<\lambda^q(t)=\lambda^q(t_0)$, where $\lambda^q(t_0)$ is the initial maximum variance of the spatial distribution of the agents.
\end{theoremEnd}
\begin{proofEnd}
By the definition  of PCA \cite{Jollifi02}, $\lambda^n=\text{argmin}_{\bm{u}} \bm{u}^\intercal\bm{C}\bm{u} $ is the smallest variance, and $\lambda^q=\text{argmax}_{\bm{u}} \bm{u}^\intercal\bm{C}\bm{u} $ is the largest variance. This implies that by definition $\lambda^n(t)\leq\lambda^q(t)$. 
What remains is to show that $\dot{\lambda}^q=0$.
Taking the time derivative of $\lambda^q=\bm{q}^\intercal\bm{C}\bm{q}
=\sum_{i=1}^M\langle \bm{r}_i-\bm{r}_c,\bm{q}\rangle^2$, we obtain
\begin{align}
&\dot{\lambda}^q
=2\sum_{i=1}^M\langle \bm{r}_i-\bm{r}_c,\bm{q}\rangle(\langle \dot{\bm{r}}_i-\dot{\bm{r}}_c,\bm{q}\rangle+\langle \bm{r}_i-\bm{r}_c,\dot{\bm{q}}\rangle)
\end{align}
But, for a complete graph $\langle \dot{\bm{r}}_i-\dot{\bm{r}}_c,\bm{q}\rangle=k_1(z_i^d-z_a^d)\langle \bm{n},\bm{q}\rangle=0$, where   $z_a=\frac{1}{M}\sum_{i=1}^M z_i$ is the average field measurement. Since from
(\ref{eq:qdot6completelinearizedx}) $\langle\bm{q} , \dot{\bm{q}}\rangle=0$, then
\begin{align}
    &\dot{\lambda}^q
=2\sum_{i=1}^M\langle \bm{r}_i-\bm{r}_c,\bm{q}\rangle\langle \bm{r}_i-\bm{r}_c,\dot{\bm{q}}\rangle
=2\lambda^q\langle\bm{q} , \dot{\bm{q}}\rangle=0.
\end{align}
Similarly, taking the time derivative of $\lambda^n=\bm{n}^\intercal\bm{C}\bm{n}
=\sum_{i=1}^M\langle \bm{r}_i-\bm{r}_c,\bm{n}\rangle^2$, 
and using $\dot{\bm{r}}_i-\dot{\bm{r}}_c=k_1(z_i-z_a)\bm{n}$, we obtain $\langle \dot{\bm{r}}_i-\dot{\bm{r}}_c,\bm{n}\rangle=k_1(z_i-z_a)\langle \bm{n},\bm{n}\rangle=k_1(z_i-z_a)$. 
Additionally, since from (\ref{eq:ndot6})  $\langle\bm{n} , \dot{\bm{n}}\rangle=0$, we obtain $\sum_{i=1}^M\langle \bm{r}_i-\bm{r}_c,\bm{n}\rangle\langle \bm{r}_i-\bm{r}_c,\dot{\bm{n}}\rangle=2\langle\bm{C}\bm{n} , \dot{\bm{n}}\rangle=2\lambda^n\langle\bm{n} , \dot{\bm{n}}\rangle=0$. Therefore
$\dot{\lambda}^n=2k_1\sum_{i=1}^M(z_i-z_a)\langle \bm{r}_i-\bm{r}_c,\bm{n}\rangle
$. 
Additionally, using (\ref{eq:field2}), $z_a=
\frac{1}{M}\sum_{k=1}^M z_k=z_c+\frac{1}{M}\sum_{k=1}^M\langle \bm{r}_k-\bm{r}_c,\nabla z_c\rangle+\frac{1}{M}\sum_{k=1}^M\nu_k =
z_c+\frac{1}{M}\sum_{k=1}^M\nu_k$, where we used the fact $\sum_{k=1}^M\langle \bm{r}_k-\bm{r}_c,\nabla z_c\rangle=0$. Hence
\begin{align}
&\dot{\lambda}^n=2k_1\sum_{i=1}^M(z_i-z_c)\langle \bm{r}_i-\bm{r}_c,\bm{n}\rangle,
\end{align}
where we used the fact that $\sum_{i=1}^M\sum_{k=1}^M\nu_k\langle \bm{r}_i-\bm{r}_c,\bm{n}\rangle=\sum_{k=1}^M\nu_k\sum_{i=1}^M\langle \bm{r}_i-\bm{r}_c,\bm{n}\rangle=0$.
Finally, using (\ref{eq:field2}), we can further write 
\begin{align}\label{lamdandotall}
&\dot{\lambda}^n=2k_1\lambda^n\|\nabla z_c\|\langle \bm{N}_c,\bm{n}\rangle+2k_1\sum_{i=1}^M\nu_i\langle \bm{r}_i-\bm{r}_c,\bm{n}\rangle.
\end{align}
Note that, from (\ref{lamdandotall}), $\lambda^n$ will increase or decrease, hence the shape will stretch or shrink along the $\bm{n}$ direction, depending on the signs of $\langle\bm{N}_c,\bm{n}\rangle$  and $\sum_{i=1}^M\nu_i\langle \bm{r}_i-\bm{r}_c,\bm{n}\rangle$. However, if it increases, it will do so only up to $\lambda^n=\lambda^q$. At this point, since $\lambda^n=\text{argmin}_{\bm{u}} \bm{u}^\intercal\bm{C}\bm{u}$ and $\lambda^q=\text{argmax}_{\bm{u}} \bm{u}^\intercal\bm{C}\bm{u}$, then  the PCA flow will interchange $\bm{n}$ and $\bm{q}$ and hence the swarm performs a turn of at most $90^\circ$. Consequently, $\lambda^q(t)\leq\lambda^q(t_0)$. 
\end{proofEnd}
Since $\lambda^q$ is the largest variance of the swarm, then
 \textbf{Lemma~\ref{'lambdaqdot is zero'}} implies that variance of the swarm is bounded by the initial variance $\lambda^q(t_0)$ . Consequently, the connectivity of the graph is ensured even without a formation controller.

\section{ Convergence Analysis} \label{Convergence}
In this section, we first obtain conditions under which  we  prove that (A) for  complete graphs, the SUSD direction $\bm{n}$ converges to the negative gradient direction $-\bm{N}_{c}$ for both source seeking and level curve tracking, and (B)
for  incomplete graphs, the SUSD direction $\bm{n}_i$ of each agent converges to the negative  gradient direction $-\bm{N}_{c,i}$ for source seeking.\update{ 
Then, we provide conditions under which we prove that (C) for complete graphs, trajectories $z(\bm{r}_c)-z^d$  for   level curve tracking, or  of $z(\bm{r}_c)$ for source seeking, are ultimately bounded. Finally,  for incomplete graphs, in (D)  we provide conditions under which  trajectories of $z(\bm{r}_{i,c})$   for the source seeking are ultimately bounded.
}

Recall that the PCA flow (\ref{eq:PCAflow}) runs in the time scale $\tau$, while the  control law (\ref{eq:controllaw}) runs in the time scale $t$. That is, for each time instance $t$, each agent runs (\ref{eq:PCAflow}) for some time $\tau$. 
Let the relationships between the control time $t$, and the PCA perception time $\tau$ be  $\frac{dt}{d\tau}=\epsilon$, where $\epsilon\in(0,1)$. This implies that $\tau=\frac{t-t_0}{\epsilon}$, where $\tau_0=0$. Using this relationship, the perception and control dynamics in the singular perturbation framework are 
\begin{align}
&\dot{\bm{r}}_i=k_1(z_i-z^d)\bm{n}_i+k_2\bm{q}_i,\quad\forall i,\label{eq:sloworiginal1}\\
&\dot{\bm{n}}_i=g_1(\cdot),\label{eq:sloworiginal2}\\
&\dot{\bm{q}}_i=g_2(\cdot),\label{eq:sloworiginal3}\\
&\epsilon\dot{\hat{\bm{q}}}_i=(\bm{I}-\hat{\bm{q}}_i\hat{\bm{q}}_i^\intercal)\bm{C}_i\hat{\bm{q}}_i,\quad\forall i.\label{eq:fastoriginal}
\end{align}
where $g_1(\cdot)$ and $g_2(\cdot)$ are the general information dynamics equations given by (\ref{eq:ndot6}) and (\ref{eq:qdot6}). 
The  control dynamics  (\ref{eq:sloworiginal1})-(\ref{eq:sloworiginal3}) are viewed as a slow system whereas the perception dynamics (\ref{eq:fastoriginal}) are viewed as a fast system.


\subsection{Convergence of the SUSD Direction for  Complete Graphs for Source Seeking and Level Curve Tracking}
We view the swarm as  one body where its individuals are moving in the same direction but with different speeds depending on their field measurements. Define 
\begin{align}\label{theta}
    & \theta=1+\langle\bm{N}_c,\bm{n}\rangle,
\end{align}
where $\theta\to0$ when $\bm{n}\to -\bm{N}_c$. i.e. when the swarm speeds up or slows down in the negative direction of the field gradient.
Additionally, define 
\begin{align}\label{psi}
    & \psi=1- \langle\bm{q},\hat{\bm{q}}\rangle,
\end{align}
where $\psi\to0$ when $\hat{\bm{q}}\to \bm{q}$, i.e when the PCA perception algorithm  converges to the exact eigenvector of the covariance matrix $\bm{C}$.
We then obtain the coerced slow and fast systems
\begin{align}
    &\dot{\theta}=k_1\|\nabla z_c\|\frac{\lambda^q}{\lambda^q-\lambda^n}\theta(\theta-2)
+\delta,\label{slowsystemcomplete}\\
&\epsilon\dot{\psi}=-(\lambda^q-\lambda^n)\psi(1-\psi)(2-\psi)+\epsilon\eta,\label{fastsystemcomplete}
\end{align}
where $\delta$ is viewed as an input disturbance  due  to  the   nonlinearity of the field, and  $\eta$ represents the interconnection between the coerced slow and fast systems.
They are defined by
\begin{align}\label{noisecomplete}
    &\delta=-\frac{k_1}{\lambda^q-\lambda^n}\vartheta\langle\bm{N}_c,\bm{q}\rangle
    +\langle\bm{n},\dot{\bm{N}_c}\rangle,\notag\\
    &\eta=\pm\frac{k_1}{\lambda^q-\lambda^n}\Big(\vartheta\pm \|\nabla z_c\|\lambda^q\sqrt{\theta(2-\theta)}\Big)\sqrt{\psi(2-\psi)},
\end{align}
where $\vartheta=\sum_{k=1}^M\nu_k\langle\bm{r}_k-\bm{r}_c,\bm{q}\rangle$, and $\langle\bm{n},\dot{\bm{N}_c}\rangle=\frac{1}{\|\nabla z_c\|}\bm{n}^\intercal(\bm{I}-\bm{N}_c\bm{N}_c^\intercal)\nabla^2z_c(k_1(z_a-z_d)\bm{n}+k_2\bm{q})$, where $z_a$ is the average field measurement and $\nabla^2z_c$ is the hessian matrix of the field.

\begin{proof}\textit{of (\ref{slowsystemcomplete}) and (\ref{fastsystemcomplete}).}
To derive (\ref{slowsystemcomplete}), we take the time derivative of (\ref{theta}) and apply (\ref{eq:ndot6complete22}) of \textbf{Lemma}~\ref{lemmabodyframeboth} for $\dot{\bm{n}}$. On the other hand, we derive (\ref{fastsystemcomplete}) by the following steps. By the Chain rule, $\frac{d\hat{\bm{q}}}{d\tau}=\epsilon\frac{d\hat{\bm{q}}}{dt}$, or $\frac{d\hat{\bm{q}}}{dt}=\frac{1}{\epsilon}\frac{d\hat{\bm{q}}}{d\tau}$. Hence
\begin{align}\label{eq:ag}
&\frac{d\psi}{d\tau}=\epsilon\frac{d\psi}{dt}
=-\epsilon\langle\frac{d\bm{q}}{dt} , \hat{\bm{q}}\rangle-\langle\bm{q} , \frac{d\hat{\bm{q}}}{d\tau} \rangle.
\end{align}
From (\ref{eq:PCAflow}), we obtain 
\begin{align}\label{ag1}
    &\langle\bm{q} , \frac{d\hat{\bm{q}}}{d\tau} \rangle= (1-\psi)\Big(\lambda^q-\langle\hat{\bm{q}}, \bm{C}\hat{\bm{q}}\rangle\Big).
\end{align}
Write $\hat{\bm{q}}=\langle \hat{\bm{q}},\bm{q}\rangle\bm{q}+\langle \hat{\bm{q}},\bm{n}\rangle\bm{n}$. Hence
\begin{align}\label{qhatCqhat}
    &\langle\hat{\bm{q}}, \bm{C}\hat{\bm{q}}\rangle
    =\lambda^q(1-\psi)^2+\lambda^n\psi(2-\psi).
\end{align}
Substituting (\ref{qhatCqhat}) into (\ref{ag1}) yields
\begin{align}\label{qdqhattau}
    &\langle\bm{q} , \frac{d\hat{\bm{q}}}{d\tau} \rangle= (\lambda^q-\lambda^n)\psi(1-\psi)(2-\psi).
\end{align}
On the other hand, using (\ref{eq:qdot6completelinearizedx}), we obtain
\begin{align}\label{ag2}
&\langle\frac{d\bm{q}}{dt} , \hat{\bm{q}}\rangle=\notag\\
&\frac{k_1}{\lambda^q-\lambda^n}\Big(\|\nabla z_c\|\lambda^q\langle \bm{N}_c,\bm{q} \rangle
+\vartheta\Big)\langle\bm{n},\hat{\bm{q}}\rangle,
\end{align} 
where $\vartheta=\sum_{k=1}^M\nu_k\langle\bm{r}_k-\bm{r}_c,\bm{q}\rangle$, and $\langle\bm{n},\hat{\bm{q}}\rangle=\pm\sqrt{\psi(2-\psi)}$.
Substituting (\ref{qdqhattau}) and (\ref{ag2}) in (\ref{eq:ag}), we obtain
\begin{align}\label{eq:dynamics_of_fastsystem_tau_domain}
    &\frac{d\psi}{d\tau}=\epsilon\frac{d\psi}{dt}=-(\lambda^q-\lambda^n)\psi(1-\psi)(2-\psi)+\epsilon\eta,
\end{align}
where $\eta$ is as defined by (\ref{noisecomplete}).
\end{proof}

We first let $\epsilon=0$ in (\ref{slowsystemcomplete}) and (\ref{fastsystemcomplete}) to  analyze the stability of the resulting decoupled reduced and boundary systems (\ref{SUSDreducedComplete}) and (\ref{eq:boundary_dynamics}), respectively.
Then, we analyze the  stability of the coupled system of (\ref{slowsystemcomplete}) and (\ref{fastsystemcomplete}) 
by deriving  $\epsilon^*\in(0,1)$ such that for all $\epsilon\leq\epsilon^*$, some of the stability results of the  reduced and boundary systems, (\ref{SUSDreducedComplete}) and (\ref{eq:boundary_dynamics}), hold for the coupled system. 

\subsubsection{Stability of the Reduced System}
The coerced reduced system is given by
\begin{align}\label{SUSDreducedComplete}
&\dot{\theta}= -k_1\|\nabla z_c\|\frac{\lambda^q}{\lambda^q-\lambda^n}\theta(2-\theta)
+\delta=f(t,\theta,\delta).
\end{align}
Note that, when $\theta\in\{0,2\}$, then $\bm{n}=\pm\bm{N}_c$ which implies that $\langle\bm{N}_c,\bm{q}\rangle=0$ and $\bm{n}^\intercal(\bm{I}-\bm{N}_c\bm{N}_c^\intercal)=0$. Hence $\delta$ vanishes at the equilibria $\theta\in\{0,2\}$. Additionally observe that $\delta=0$ when $\nabla^2z_c=0$ and $\nu_k=0$, $\forall k$, i.e. when the field is  linear.

The following result describes the stability of the origin of the reduced system. 
\begin{theoremEnd}{thm}\label{'lemma_11'}
Consider the reduced system (\ref{SUSDreducedComplete}).
Suppose there exists a lower bound $\mu_1>0$ such that $\|\nabla z(\bm{r}_c)\|>\mu_1$.
Then the equilibrium $\theta=0$ of  the unforced system $f(t,\theta,0)$ is asymptotically stable in which whenever $\theta(0)\in[0,2)$, then $\theta(t)\to0$ as $t\to\infty$.
Furthermore, for an input disturbance satisfying  $|\delta|\leq k_1\epsilon_1\frac{\lambda^q}{\lambda^q-\lambda^n}\mu_1$, where $\epsilon_1\in(0,1)$, the equilibrium $\theta=0$ of forced system $f(t,\theta,\delta)$ is locally input-to-state stable. 
\end{theoremEnd}

\begin{proof}
Consider the domain $\bm{D}_1=\{\theta|\theta\in[0,2)\}$
i.e. $ \langle\bm{N}_c,\bm{n}\rangle\neq1$.  Let $V_1:\bm{D}_1\to\bm{R}$ be a Lyapunov candidate function defined by
\begin{align}\label{V1equation}
    &V_1=\frac{\theta}{2-\theta},
\end{align}
where $V_1=0$ if and only if $\theta=0$. Additionally, $V_1\to\infty$ as $\theta\to2$.
For the unforced system $f(t,\theta,0)$, we obtain
\begin{align}
    \dot{V}_1&=-2k_1\|\nabla z_c\|\frac{\lambda^q}{\lambda^q-\lambda^n}V_1\leq0.
\end{align}
Since $\dot{V}_1=0$ if and only if $\theta=0$, then the origin of the unforced system $f(t,\theta,0)$ is asymptotically stable. Additionally, $\dot{V}_1\to-\infty$ as $\theta\to2$. This along with the fact that  $V_1\to\infty$ whenever $\theta\to 2$ and $\|\nabla z_c\|>\mu_1>0$, implies that  $\bm{D}_1$ is a forward invariant set, and thus $\theta\in[0,2)$ for all $t$.

For the forced system $f(t,\theta,\delta)$, we obtain
\begin{align}\label{V1dotfinal}
    \dot{V_1}&\leq-2(1-\epsilon_1)k_1\mu_1\frac{\lambda^q}{\lambda^q-\lambda^n} V_1,\quad\forall |\theta|\geq\rho(|\delta|),
\end{align}
where $\rho(|\delta|)=1-\sqrt{1-\frac{(\lambda^q-\lambda^n)|\delta|}{k_1\epsilon_1\lambda^q\mu_1}}$ is a class $\mathcal{K}$  function in the domain $[0,k_1\epsilon_1\frac{\lambda^q}{\lambda^q-\lambda^n}\mu_1]$. Let $\alpha_1(|\theta|)=\alpha_2(|\theta|)=\frac{|\theta|}{2-|\theta|}$ which are  class \update{$\mathcal{K}$} functions that satisfy: $\alpha_1(|\theta|)\leq V_1(\theta)\leq\alpha_2(|\theta|)$\footnote{For more details about the definitions of class $\mathcal{K}$  functions, the reader is referred to \textbf{Definition 4.2} of \cite{khalil2002nonlinear}.}. 
Therefore, using \textbf{Definition 3.3} of local input-to-state stability in  \cite{dashkovskiy2011input}, and according to \textbf{Theorem 4.19} in \cite{khalil2002nonlinear}, the origin of the forced system $f(t,\theta,\delta)$ is locally input-to-state stable.
\end{proof}
\update{
In \textbf{Theorem~\ref{'lemma_11'}} we showed that the set $\{\theta|\theta\in[0,2)\}$ is forward invariant. 
The following result shows that the restricted set $\{\theta|\theta\in[0,1)\}$ is also forward invariant. This result will be required later in
Section~\ref{convergence of z}.  
\begin{corollary}\label{'corollary1'}
Consider the reduced system (\ref{SUSDreducedComplete}).
Suppose there exists a lower bound $\mu_1>0$ such that $\|\nabla z(\bm{r}_c)\|>\mu_1$. 
Then the equilibrium $\theta=0$ of  the unforced system $f(t,\theta,0)$ is asymptotically stable in which whenever $\theta(0)\in[0,1)$, then $\theta(t)\to0$ as $t\to\infty$.
Furthermore, for an input disturbance satisfying  $|\delta|\leq k_1\epsilon_1\frac{\lambda^q}{\lambda^q-\lambda^n}\mu_1$, where $\epsilon_1\in(0,1)$, the equilibrium $\theta=0$ of the forced system $f(t,\theta,\delta)$ is locally input-to-state stable. 
\end{corollary}
\begin{proof}
If we modify $V_1$ in  (\ref{V1equation}) to be $V_1=\frac{2\theta}{1-\theta}$, where  $V_1:[0,1)\to\bm{R}$, then we can show that $\dot{V}_1$ satisfies (\ref{V1dotfinal}). 
Hence, using the same argument in proving \textbf{Theorem~\ref{'lemma_11'}}, we conclude that the origin of of the forced system $f(t,\theta,\delta)$ is locally input-to-state stable and the set $\{\theta|\theta\in[0,1)\}$ is forward invariant. 
\end{proof}

}
\begin{theoremEnd}{lemma}\label{mulemma}
The assumption in Theorem~\ref{'lemma_11'} that the input disturbance satisfies  $|\delta|\leq k_1\epsilon_1\frac{\lambda^q}{\lambda^q-\lambda^n}\mu_1$  is valid whenever
 $\|\nabla z(\bm{r}_c)\|>\mu_1$ where
\begin{align}\label{mumax}
    &\mu_1=\frac{|\vartheta|+\sqrt{|\vartheta|^2+4\epsilon_1\lambda^q(\lambda^q-\lambda^n)(|z_a-z_d|+\frac{k_2}{k_1})\|\nabla^2 z_c\|}}{2\epsilon_1\lambda^q},
\end{align}
\end{theoremEnd}
in which $\vartheta$ is as defined in (\ref{noisecomplete}).
\begin{proofEnd}
Since $\rho(|\delta|)$ is a real number, then we must have
$\sqrt{1-\frac{(\lambda^q-\lambda^n)|\delta|}{k_1\epsilon_1\lambda^q\mu_1}}\geq0$ which implies that $\|\nabla z_c\|>\mu_1\geq\frac{(\lambda^q-\lambda^n)|\delta|}{k_1\epsilon_1\lambda^q}$. But $|\delta|\leq|\delta_1|+|\delta_2|$ $\leq\frac{k_1}{\lambda^q-\lambda^n}|\vartheta|+\frac{k_1|z_a-z_d|+k_2}{\|\nabla z_c\|}\|\nabla^2 z_c\|$, where $\vartheta=\sum_k\nu_k\langle\bm{r}_k-\bm{r}_c , \bm{q}\rangle$ and $\nabla^2 z_c$ is the Hessian matrix. Then  we must have $\|\nabla z_c\|>\frac{1}{\epsilon_1\lambda^q}|\vartheta|+\frac{(\lambda^q-\lambda^n)(k_1|z_a-z_d|+k_2)}{k_1\epsilon_1\lambda^q\|\nabla z_c\|}\|\nabla^2 z_c\|$. Solving this inequality yields the desired result (\ref{mumax}).
\end{proofEnd}

Observe that \textbf{Theorem~\ref{'lemma_11'}} implies that wherever the swarm is in a landscape where $\|\nabla z(\bm{r}_c)\|>\mu_1$, then the SUSD direction $\bm{n}$ follows the negative gradient direction $-\bm{N}$. Since according to \textbf{Assumption~\ref{fieldassumption}} the field has a unique minimum, then the bound $\mu_1$ defines a neighborhood around the source location $\bm{r}_0$ where the magnitude of the gradient $\|\nabla z(\bm{r}_c)\|$ is dominated by the higher-order terms. Once the swarm is inside this neighborhood, then $\bm{n}$ may not track  $-\bm{N}$. Without a termination policy in the Algorithm, the swarm may pass the source. Fortunately,  \textbf{Lemma~\ref{'lambdaqdot is zero'}} shows that the swarm is guaranteed to switch between $\bm{n}$ and $\bm{q}$ and hence the swarm steers back to the set  $\|\nabla z(\bm{r}_c)\|>\mu_1$. 
\begin{remark}
Note that $\mu_1$ is a sufficient bound, and if we consider only source seeking, then we simply  substitute $k_2=0$ and $z_d=0$ in (\ref{mumax}). Intuitively, since $\vartheta=\mathcal{O}(\|\bm{r}_k-\bm{r}_c\|^2)$, then $\mu_1$ decreases as the spatial size of the swarm shrinks.
Moreover, when the swarm is more spatially distributed then $(\lambda^q-\lambda^n)$ decreases which reduce the effect of the Hessian term.
\end{remark}

\subsubsection{Stability of the Boundary System}\label{boundarycompletesection}

By setting $\epsilon=0$ in (\ref{eq:dynamics_of_fastsystem_tau_domain}),  we obtain the boundary system
\begin{align}\label{eq:boundary_dynamics}
&\frac{d\psi}{d\tau}=-(\lambda^q-\lambda^n)\psi(1-\psi)(2-\psi).
\end{align}
Observe that in (\ref{eq:boundary_dynamics}),  $\lambda^q$ and $\lambda^n$ are constants with respect to the time scale $\tau$. 
Additionally, system (\ref{eq:boundary_dynamics}) is at equilibrium when $\psi\in\{0,1,2\}$. The desired equilibrium $\psi=0$ corresponds to  $\hat{\bm{q}}=\bm{q}$, and the undesired  $\psi=1$ and $\psi=2$ correspond to   $\hat{\bm{q}}=\pm\bm{n}$ and $\hat{\bm{q}}=-\bm{q}$, respectively.
We have the following result for origin of the boundary system
\begin{theoremEnd}{thm}\label{'lemma_epsi'}
The origin of the boundary system (\ref{eq:boundary_dynamics}) is \update{asymptotically} stable uniformly in $\lambda^q$ and $\lambda^n$, in  which whenever at $\tau=0$, $\psi(0)\in[0,1)$, then $\psi\to0$ as $\tau\to\infty$.
\end{theoremEnd}
\begin{proof}
Let $\bm{D}_2=\{\psi\in \mathbb{R}| \psi\in[0,1)\}$   which is equivalent to  $0<\langle \bm{q},\hat{\bm{q}}\rangle\leq1$. Then let $V_2(\psi):\bm{D}_2\to \mathbb{R}$ be a Lyapunov candidate function defined by
\begin{align}\label{V2equation}
    &V_2=\frac{\psi}{1-\psi}
\end{align}
where $V_2\geq0$ and $V_2=0$ if and only if $\psi=0$. Furthermore, $V_2\to\infty$ as $\psi\to 1$.
Using (\ref{eq:boundary_dynamics}), we
obtain
\begin{align}\label{V2dot}
&\frac{dV_2}{d\tau}=-(\lambda^q-\lambda^n)(2-\psi)V_2
\leq0,
\end{align}
 where in $\bm{D}_2$, 
 $\frac{dV_2}{d\tau}=0$  if and only if $\psi=0$. Furthermore, since from \textbf{Assumption}~\ref{shape assumption} $(\lambda^q-\lambda^n)\neq0$, then  $\frac{dV_2}{d\tau}\to-\infty$ as $\psi\to1$. This along with the fact that  $V_2\to\infty$ whenever $\psi\to 1$ implies that  $\bm{D}_2$ is a forward invariant set, and thus $\psi(\tau)\in[0,1)$ for all $\tau$. Let $W_1(\psi)=W_2(\psi)=V_2$ which implies that $W_1(\psi)\leq V_2\leq W_2(\psi)$. 
 Consequently, according to \textbf{Theorem 4.9} in \cite{khalil2002nonlinear}, we can conclude that the
 equilibrium $\psi=0$ of the boundary system (\ref{eq:boundary_dynamics})  is asymptotically stable, uniformly in $\lambda^q$ and $\lambda^n$. 
\end{proof}

\subsubsection{Stability of the Unforced Coupled System}\label{'coupledsystemcomplete'}
Substituting for $\delta=0$ in (\ref{slowsystemcomplete}) and (\ref{fastsystemcomplete}), we obtain the unforced coupled system
\begin{align}
    &\dot{\theta}=k_1\|\nabla z_c\|\frac{\lambda^q}{\lambda^q-\lambda^n}\theta(\theta-2)\triangleq f(t,\theta(t),0,0)
,\label{slowsystemcompleteuf}\\
&\epsilon\dot{\psi}=-(\lambda^q-\lambda^n)\psi(1-\psi)(2-\psi)+\epsilon\eta\triangleq g(t,\theta(t),\psi(t),\epsilon),\label{fastsystemcompleteuf}
\end{align}
where $\eta=\pm k_1\|\nabla z_c\|\frac{\lambda^q}{\lambda^q-\lambda^n}\sqrt{\theta(2-\theta)}\sqrt{\psi(2-\psi)}$.

\begin{theoremEnd}{thm}\label{maintheorem1}
 Consider the coupled system given by (\ref{slowsystemcompleteuf}) and (\ref{fastsystemcompleteuf}). 
 Assume that $0<\mu_1\leq\|\nabla z_c\|\leq\mu_2<\infty$ and $0<\chi_1\leq\lambda^q-\lambda^n\leq\chi_2<\infty$ where $\mu_1$, $\mu_2$, $\chi_1$ and $\chi_2$ are constants.
Furthermore, assume that $\epsilon<\epsilon_d$ where 
\begin{align}\label{espsd}
     &\epsilon_d=\frac{2(1-d)\mu_1\chi_1^3}{d k_1\mu_2^2\chi_2^2},
 \end{align}
in which $d\in(0,1)$ is a constant. 
Then the origin $(\theta,\psi)=(0,0)$ is uniformly asymptotically stable in which whenever $\theta(0)\in[0,2)$ and $\psi(0)\in[0,1)$, then $(\theta(t),\psi(t))\to(0,0)$ as $t\to\infty$. 
\end{theoremEnd}
\begin{proof}
Consider $(t,\theta,\psi,\epsilon)\in[t_0,\infty)\times [0,2)\times [0,1)\times [0,\epsilon_0]$. Then, the origin $(\theta,\psi)=(0,0)$ is the unique equilibrium of $0=f(t,0,0,\epsilon)$ and $0=g(t,0,0,\epsilon)$. Moreover, $0=g(t,\theta,\psi,0)$ has a unique root $z=h(t,\theta)=0$.

In the proof of \textbf{Theorem}~\ref{'lemma_11'}, we showed that the positive definite $V_1(t,\theta)=\frac{\theta}{2-\theta}$ satisfies (i): $\alpha_1(|\theta|)\leq V_1(t,\theta)\leq\alpha_2(|\theta|)$ where $\alpha_1(|\theta|)=\alpha_2(|\theta|)=\frac{|\theta|}{2-|\theta|}$ are class $\mathcal{K}$ functions, and, since $\|\nabla z_c\|\geq\mu_1$ and $\frac{\lambda^q}{\lambda^q-\lambda^n}\geq1$, (ii): $\frac{\partial V_1}{\partial t}+\frac{\partial V_1}{\partial \theta}f(t,\theta,0,0)\leq
-\varrho_1U_1^2(\theta)$ where $\varrho_1=2k_1\mu_1$ is a constant, and $U_1(\theta)=\sqrt{\frac{\theta}{2-\theta}}$ is a continuous scalar function that vanishes only when $\theta=0$.

Similarly, in the proof of  \textbf{Theorem}~\ref{'lemma_epsi'}, we showed that the positive definite $V_2(t,\psi)=\frac{\psi}{1-\psi}$ satisfies (iii) $W_1(\psi)\leq V_2(t,\psi)\leq W_2(\psi)$ where $W_1(\psi)=W_2(\psi)=\frac{\psi}{1-\psi}$ are class  $\mathcal{K}$ functions, and, since $(\lambda^q-\lambda^n)\geq\chi_1$, (iv):
$\frac{\partial V_2}{\partial \psi}g(t,\theta,\psi,0)\leq-\varrho_2U_2^2(\psi)$, where $\varrho_2=\chi_1$ and $U_2(\psi)=\sqrt{\frac{\psi(2-\psi)}{1-\psi}}$ is  a continuous scalar function that vanishes only when $\psi=0$.

For the interconnected system, we have (v): $\frac{\partial V_1}{\partial \theta}[f(t,\theta,\psi,\epsilon)-f(t,\theta,h(t,x),0)]=\frac{\partial V}{\partial \theta}[f(t,\theta,0,0)-f(t,\theta,0,0)]=0$ where we used the fact that $f$ is independent of $\psi$ and $\epsilon$. Similarly, since $V_2$ does explicitly depend on $t$ and $\theta$, we have (vi): $ \frac{\partial V_2}{\partial t}+\frac{\partial V_2}{\partial \theta}f(t,\theta,\psi,\epsilon)=0$. Finally, we derive (vii): $\frac{\partial V_2}{\partial \psi}[g(t,\theta,\psi,\epsilon)-g(t,\theta,\psi,0)]=\epsilon\eta\leq\epsilon \varrho_3U_1(\theta)U_2(\psi)$, where $\varrho_3=2 k_1\mu_2\frac{\chi_2}{\chi_1}$. 

Through (i) to (vii), we satisfy all the assumptions required by \textbf{Theorem 5.1} in \cite{Kokotovi1986SingularPM}. 
Hence, according to \textbf{Theorem 5.1} in \cite{Kokotovi1986SingularPM},  for every $d\in(0,1)$, $v(t,\theta,\psi)=(1-d)V_1(t,\theta)+dV_2(t,\psi)$ is a Lyapunov function for all $\epsilon<\epsilon_d$ where  $\epsilon_d$ is given by\footnote{We derived $\epsilon_d$ by using (5.12) in \cite{Kokotovi1986SingularPM} using all the corresponding coefficients derived in (i) through (vii).} (\ref{espsd}). This implies that 
the origin $(\theta,\psi)=(0,0)$ is a uniformly asymptotically stable equilibrium of the singularly perturbed system  given by (\ref{slowsystemcompleteuf}) and (\ref{fastsystemcompleteuf}) for all $\epsilon\in (0,\epsilon_0)$.  
\end{proof}

\subsection{Convergence of the SUSD Direction for Source Seeking under   Incomplete Graphs }

Similar to the complete graph case, we first define 
\begin{align}\label{theta_i}
    & \theta_i=1+\langle\bm{N}_{c,i},\bm{n}_i\rangle,
\end{align}
where  $\theta_i\to0$ when $\bm{n}_i\to -\bm{N}_{c,i}$. i.e. when $\bm{n}_i$ converges to the  negative direction of the local field gradient, $\bm{N}_{c,i}=\frac{\nabla z_{c,i}}{\|\nabla z_{c,i}\|}$.
Additionally, define 
\begin{align}\label{psi_i}
    & \psi_i=1- \langle\bm{q}_i,\hat{\bm{q}}_i\rangle,
\end{align}
where $\psi_i\to0$ when $\hat{\bm{q}}_i\to \bm{q}_i$ $\forall i$, i.e when all the local PCA perception algorithms  converge to the exact eigenvectors of the local covariance matrices $\bm{C}_i$.
Using  \textbf{Corollary}~\ref{corolaryincomplete}, and similar to the procedure of deriving (\ref{slowsystemcomplete}), (\ref{fastsystemcomplete}), and (\ref{noisecomplete}),  we obtain the coerced slow and fast systems for incomplete graphs
\begin{align}
    &\dot{\theta}_i=k_1\|\nabla z_{c,i}\| \frac{\lambda_i^q}{\lambda_i^q-\lambda_i^n}\theta_i(\theta_i-2)+\delta_i,i=1,\cdots, M,\label{slowsystemincomplete}\\
&\epsilon\dot{\psi}_i=-(\lambda_i^q-\lambda_i^n)\psi_i(1-\psi_i)(2-\psi_i)+\epsilon \eta_i\label{fastsystemincomplete}
\end{align}
where $\delta_i$  and  $\eta_i$ are defined by
\begin{align}\label{noisetermsincomplete}
    &\delta_i=-\langle\bm{N}_{c,i},\bm{q}_i\rangle\mathcal{E}_i-\frac{k_1}{\lambda_i^q-\lambda_i^n}\vartheta_i\langle\bm{N}_{c,i},\bm{q}_i\rangle
       +\langle\bm{n}_i,\dot{\bm{N}}_{c,i}\rangle,\notag\\
    &\eta_i=\pm\frac{k_1}{\lambda_i^q-\lambda_i^n}\Big(\vartheta_i+\mathcal{E}_i\pm\|\nabla z_{c,i}\|\lambda_i^q\sqrt{\theta_i(2-\theta_i)}
\Big)\cdot\notag\\
    &\sqrt{\psi_i(2-\psi_i)},
\end{align}
where $\mathcal{E}_i$, as defined in \textbf{Corollary}~\ref{corolaryincomplete},  accounts for the mismatch between the local PCA components. Additionally,
$\vartheta_i=\sum_{k\in\mathcal{H}_i}\nu_k \langle\bm{r}_k-\bm{r}_{c,i},\bm{q}_i\rangle$  and $\langle\bm{n}_i,\dot{\bm{N}}_{c,i}\rangle=(k_1/\|\nabla z_{c,i}\|)\bm{n}_i^\intercal(\bm{I}-\bm{N}_{c,i}\bm{N}_{c,i}^\intercal)\nabla^2 z_i(1/M_i)\sum_{k\in\mathcal{H}_i}z_k\bm{n}_k$,  where $\nabla^2z_i$ is the local hessian matrix of the field around $\bm{r}_{c,i}$.

\subsubsection{The Reduced System}
The coerced reduced system is given by
\begin{align}\label{ReducedSystemnonComplete}
&\dot{\theta}_i=k_1\|\nabla z_{c,i}\| \frac{\lambda_i^q}{\lambda_i^q-\lambda_i^n}\theta_i(\theta_i-2)+\delta_i.
\end{align}
Observe that, when $\theta_i=0,2$, then $\bm{n}_i=\pm\bm{N}_{c,i}$ which implies that $\langle\bm{N}_{c,i},\bm{q}_i\rangle=0$ and $\bm{n}_i^\intercal(\bm{I}-\bm{N}_{c,i}\bm{N}_{c,i}^\intercal)=0$. Hence, $\delta_i$ in (\ref{noisetermsincomplete}) vanishes 
 at the equilibria of (\ref{SUSDreducedComplete}). The term, $\mathcal{E}_i$, is determined by the graph structure,  which  decreases as the connectivity of the static graph increases.
 
Define  $\bm{\theta}=[\theta_1,\cdots,\theta_M]^\intercal$, and $\bm{\delta}=[\delta_1,\cdots,\delta_M]^\intercal$. Let $\dot{\theta}=[f_1(\theta_1,\delta_1),\cdots,f_M(\theta_M,\delta_M)]^\intercal$ where  $f_i(t,\theta_i,\delta_i)$ is as defined by (\ref{ReducedSystemnonComplete}).

\begin{theoremEnd}{proposition}\label{'lemma_z_noncomplete'}
Consider the reduced system (\ref{ReducedSystemnonComplete}). 
Suppose there exists a lower bound $\mu_2>0$ such that $\|\nabla z(\bm{r}_{c,i})\|>\mu_2$, $\forall i$. 
Then the equilibrium $\bm{\theta}=\bm{0}$ of  the unforced system $f(t,\bm{\theta},\bm{0})$ is asymptotically stable in which whenever $\forall i$, $\theta_i(0)\in[0,2-\varsigma]$ where $\varsigma\in(0,1)$, then $\theta_i(t)\to0$ as $t\to\infty$.
Furthermore, for an input disturbance \update{ satisfying $\|\bm{\delta}\|\leq k_1\epsilon_2\barbelow{\lambda}\mu_2$, } where $\barbelow{\lambda}=\min_i\frac{\lambda_i^q}{\lambda_i^q-\lambda_i^n}$ ,  $\epsilon_2\in(0,1)$, then the origin of forced system $f(t,\bm{\theta},\bm{\delta})$ is locally input-to-state stable.
\end{theoremEnd}
\begin{proof}
 Consider the domain \update{$\bm{D}_3=\{\bm{\theta}|\quad \|\bm{\theta}\|\in[0,\sqrt{M}(2-\varsigma)]\}\equiv\{\bm{\theta}|\forall i,\quad \theta_i\in[0,2-\varsigma]\}$}, where $\varsigma\in(0,1)$.  Let $V_3:\bm{D}_3\to\bm{R}$ be a  Lyapunov candidate function defined by
\begin{align}\label{V3equation}
    &V_3=\sum_{i=1}^M\frac{\theta_i}{2-\theta_i},
\end{align}
where  $V_3=0$ if and only if $\bm{\theta}=\bm{0}$. Additionally, $V_3\to\infty$ whenever any or all $\theta_i\to2$. For the unforced system $f(t,\bm{\theta},0)$,  we obtain
\begin{align}
    &\dot{V}_3=-2k_1\sum_{i=1}^M \|\nabla z_{c,i}\|\frac{\lambda_i^q}{\lambda_i^q-\lambda_i^n}\frac{\theta_i}{2-\theta_i}
    \leq0,
\end{align}
where $\dot{V}_3=0$ if and only if $\bm{\theta}=\bm{0}$.  Then the origin of the unforced system $f(t,\bm{\theta},\bm{0})$ is asymptotically stable. Additionally, $\dot{V}_3\to-\infty$ whenever any or all $\theta_i\to2$. This along with the fact that  $V_3\to\infty$ whenever any or all $\theta_i\to2$, implies that  $\bm{D}_3$ is a forward invariant set and thus trajectories start inside it will never go outside it.

For the forced system $f(t,\bm{\theta},\bm{\delta})$,
let $\epsilon_2\in(0,1)$ be a constant, then we obtain 
\begin{align}\label{V1dotnoncomplete}
    &\dot{V}_3\leq-2k_1(1-\epsilon_2)\barbelow{\lambda}\mu_2 V_3,\quad
    \forall \|\bm{\theta}\|>\rho(\|\bm{\delta}\|),
\end{align}
where   $\barbelow{\lambda}=\min_i\frac{\lambda_i^q}{\lambda_i^q-\lambda_i^n}$ and 
\update{
\begin{align}\label{rho1functionnoncomplete}
    &   \rho(\|\bm{\delta}\|)=\sqrt{M}-\sqrt{M-\frac{M}{k_1\epsilon_2\barbelow{\lambda}\mu_2}\|\bm{\delta}\|}
\end{align}
}
is  a class $\mathcal{K}$  function in the domain \update{ $\|\bm{\delta}\|\in [0,k_1\epsilon_2\barbelow{\lambda}\mu_2]$. Since it is assumed that $\|\bm{\delta}\|\leq k_1\epsilon_2\barbelow{\lambda}\mu_2$, then $\|\bm{\delta}\|\leq\sqrt{M}$, and hence the set $\|\bm{\theta}\|\in[\rho(\|\delta\|),(2-\|\bm{\theta}\|)\sqrt{M})$ is not empty.
Let $\alpha_3(\|\bm{\theta}\|)=\frac{1}{2-\varsigma}\|\bm{\theta}\|$, and 
$\alpha_4(\|\bm{\theta}\|)=\frac{\sqrt{M}}{\varsigma}\|\bm{\theta}\|$ which, in the domain $\bm{D}_3$, are  class $\mathcal{K}$ functions that satisfy: $\alpha_3(\|\bm{\theta}\|)\leq V_3(\bm{\theta})\leq\alpha_4(\|\bm{\theta}\|)$.}
Therefore, using \textbf{Definition 3.3} of local input-to-state stability in  \cite{dashkovskiy2011input}, and according to \textbf{Theorem 4.19} in \cite{khalil2002nonlinear}, the origin of the forced system $f(t,\bm{\theta},\bm{\delta})$ is locally input-to-state stable.
\end{proof}
Using a similar procedure of deriving (\ref{mumax}), we can obtain the  sufficient bound \update{ $\mu_2=\min_i\mu_{2,i}$, where
\begin{align}\label{mumaxincomplete}
    &\mu_{2,i}=\notag\\
    &\frac{|\vartheta_i+\mathcal{E}_i|+\sqrt{|\vartheta_i+\mathcal{E}_i|^2+4\epsilon_2z_{a,i}\lambda_i^q(\lambda_i^q-\lambda_i^n)\|\nabla^2 z_{i,c}\|}}{2\epsilon_2\lambda_i^q},
\end{align}
}
in which $\vartheta_i$ and $\mathcal{E}_i$ are as defined in (\ref{noisetermsincomplete}).

\update{
In \textbf{Proposition~\ref{'lemma_z_noncomplete'}} we showed that the set $\{\bm{\theta}|\forall i,\quad\theta_i\in[0,2)\}$ is forward invariant. 
The following result shows that the restricted set $\{\bm{\theta}|\forall i,\quad\theta_i\in[0,1)\}$ is also forward invariant. This result will be required later in
Section~\ref{convergence of zincomplete}.  
\begin{corollary}\label{'corollary1_noncomplete'}
Consider the reduced system (\ref{ReducedSystemnonComplete}). 
Suppose there exists a lower bound $\mu_2>0$ such that $\|\nabla z(\bm{r}_{c,i})\|>\mu_2$, $\forall i$. 
Then the equilibrium $\bm{\theta}=\bm{0}$ of  the unforced system $f(t,\bm{\theta},\bm{0})$ is asymptotically stable in which whenever $\forall i$, $\theta_i(0)\in[0,1)$, then $\theta_i(t)\to0$ as $t\to\infty$.
Furthermore, for an input disturbance \update{ satisfying $\|\bm{\delta}\|\leq k_1\epsilon_2\barbelow{\lambda}\mu_2$, } where $\barbelow{\lambda}=\min_i\frac{\lambda_i^q}{\lambda_i^q-\lambda_i^n}$ ,  $\epsilon_2\in(0,1)$, then the origin of forced system $f(t,\bm{\theta},\bm{\delta})$ is locally input-to-state stable.
\end{corollary}
\begin{proof}
If we modify $V_1$ in  (\ref{V3equation}) to be $V_3=\sum_{i=1}^M\frac{2\theta_i}{1-\theta_i}$, where  $V_3:[0,1)\to\bm{R}$, then we can show that $\dot{V}_3$ satisfies (\ref{V1dotnoncomplete}). 
Hence, using the same argument in proving \textbf{Proposition~\ref{'lemma_z_noncomplete'}}, we conclude that the origin of of the forced system $f(t,\theta,\delta)$ is locally input-to-state stable and the set $\{\bm{\theta}|\forall i,\quad\theta_i\in[0,1)\}$ is forward invariant. 
\end{proof}

}
\subsubsection{The Boundary System}
Setting $\epsilon=0$ in (\ref{fastsystemincomplete}) with $\frac{d\psi_i}{d\tau}=\epsilon\frac{d\psi_i}{dt}$,  we obtain the boundary system
\begin{align}\label{eq:boundary_dynamics_non}
&\frac{d\psi_i}{d\tau}=-(\lambda_i^q-\lambda_i^n)\psi_i(1-\psi_i)(2-\psi_i),\quad \forall i.
\end{align}
\begin{theoremEnd}{proposition}\label{'lemma_epsinon'}
The origin of the boundary system (\ref{eq:boundary_dynamics_non}) is \update{asymptotically} stable uniformly in  all $\lambda_i^q$ and $\lambda_i^n$, in  which whenever at $\tau=0$, all $\psi_i(0)\in[0,1)$, then all $\psi_i\to0$ as $\tau\to\infty$.
\end{theoremEnd}
\begin{proof}
Let $\bm{D}_4=\{\psi|\forall i,\psi_i\in[0,1)\}$  
where $\psi_i=1-\langle \bm{q}_i,\hat{\bm{q}}_i\rangle$.  Then let $V_4(\psi):\bm{D}_2\to \mathbb{R}$ be a Lyapunov candidate function defined by
\begin{align}\label{V4equation}
    &V_4=\sum_{i=1}^M\frac{\psi_i}{1-\psi_i},
\end{align}
where $V_4\geq0$ and $V_4=0$ if and only if $\psi_i=0$, $\forall i$. Furthermore, $V_4\to\infty$ as any $\psi_i\to 1$.
Using (\ref{eq:boundary_dynamics_non}), we obtain
obtain
\begin{align}\label{V2dotnon}
&\frac{dV_4}{d\tau}
=-\sum_{i=1}^M\frac{(\lambda_i^q-\lambda_i^n)\psi_i(2-\psi_i)}{(1-\psi_i)}
\leq0.
\end{align}
 where in $\bm{D}_4$
 $\frac{dV_4}{d\tau}=0$  if and only if $\psi_i=0$, $\forall i$. Furthermore, since from \textbf{Assumption}~\ref{shape assumption} $(\lambda_i^q-\lambda_i^n)\neq0$, $\forall i$, then  $\frac{dV_2}{d\tau}\to-\infty$ whenever any $\psi_i\to1$. This along with the fact that  $V_2\to\infty$ whenever any $\psi_i\to 1$ implies that  $\bm{D}_2$ is a forward invariant set, and thus all $\psi_i(\tau)\in[0,1)$ for all $\tau$.
 Let $\bm{\psi}=[\psi_1,\cdots,\psi_M]$ and  $W_3(\bm{\psi})=W_4(\bm{\psi})=V_4$ which implies that $W_3(\bm{\psi})\leq V_4\leq W4(\bm{\psi})$. 
 Consequently, according to \textbf{Theorem 4.9} in \cite{khalil2002nonlinear}, we can conclude that the
 equilibrium $\bm{\psi}=0$ of the boundary system (\ref{eq:boundary_dynamics_non})  is asymptotically stable, uniformly in all $\lambda_i^q$ and $\lambda_i^n$. 
\end{proof}
\subsubsection{The Coupled System}
Let   $\bm{\theta}=[\theta_1,\cdots,\theta_M]^\intercal$, $\bm{\psi}=[\psi_1,\cdots,\psi_M]^\intercal$, and $\bm{\delta}=[\delta_1,\cdots,\delta_M]^\intercal$.
Define the coupled system $\begin{bmatrix}\dot{\bm{\theta}} \quad \epsilon\dot{\bm{\psi}}\end{bmatrix}^\intercal=h(t,\epsilon,\bm{\theta},\bm{\psi},\bm{\delta})$  where  $\dot{\theta}_i$ and $\dot{\psi}_i$ are as defined by (\ref{slowsystemincomplete}) and (\ref{fastsystemincomplete}), respectively. 

\begin{theoremEnd}{proposition}\label{'coupledsystemnoncomplete'}
Consider the coupled system given by (\ref{slowsystemincomplete}) and (\ref{fastsystemincomplete}). \update{Suppose that the shape covariance eigenvalues satisfy $\forall i$, $0<\chi_1\leq\lambda_i^q-\lambda_i^n\leq\chi_2<\infty$, where $\chi_1$ and $\chi_2$ are constants.} 
Consider the neighborhood \update{ $\mathcal{B}=\{\bm{r}_{c,i}|0<\mu_2\leq|\|\nabla z(\bm{r}_{c,i})\|\leq\bar{\mu}_{2}<\infty\}$ $\forall i$, where $\mu_2$ is a lower bound given by (\ref{mumaxincomplete}) and $\bar{\mu}_2$ is a finite upper bound.} 
Furthermore, assume that $\epsilon<\epsilon^*$ where 
\update{
\begin{align}\label{espstarnoncomoplete}
     &\epsilon^*=\frac{2(1-d)\mu_2\chi_1^3}{dk_1\bar{\mu}_2^2\ell^3\chi_2^2},
 \end{align}
 }
in which $d\in(0,1)$ and $ \ell\geq1$ are constants.
Then the origin of the unforced  system $h(t,\epsilon,\bm{\theta},\bm{\psi},\bm{0})$ is uniformly asymptotically stable in which whenever, $\forall i$,  $\theta_i(0)\in[0,2)$ and $\psi_i(0)\in[0,1-\frac{1}{ \ell})$, then $(\bm{\theta}(t),\bm{\psi}(t))\to(\bm{0},\bm{0})$ as $t\to\infty$. 
\end{theoremEnd}
\begin{proof}
Consider the domain $\bm{D}=\bm{D}_3\cup \bm{D}_4=$ where $\bm{D}_3=\{\bm{\theta}|\forall i, \theta_i\in[0,2)\}$ and $\bm{D}_4=\{\psi|\forall i,\psi_i\in[0,1-\frac{1}{\ell})\}$.
Let $\mathcal{V}_2:D\to\mathbb{R}$  be a Lyapunov candidate function for the coupled system and defined as $\mathcal{V}_2=(1-d)V_3+dV_4$ where $d\in(0,1)$, and $V_3$ and $V_4$ are as defined by (\ref{V3equation}) and (\ref{V4equation}), respectively.
Since $\epsilon\neq0$, then using (\ref{fastsystemincomplete}), we obtain
\begin{align}\label{V2dotcoupled_non}
    &\dot{V}_4=-\frac{1}{\epsilon}L_4+Q_4,
\end{align}
where $L_4=\sum_{i=1}^M\frac{(\lambda_i^q-\lambda_i^n)\psi_i(2-\psi_i)}{(1-\psi_i)}$ is a continuous positive definite function in the domain $\bm{D}_4$, and $Q_4=\sum_{i=1}^M\frac{\eta_i}{(1-\psi_i)^2}$ is  an indefinite function.
Following similar procedure of proving Theorem~\ref{maintheorem1}, and 
after many manipulations, we show $Q_4\leq h\sqrt{V_3}\sqrt{\bar{V}_4}$ where
\update{$h=2k_1\sqrt{\ell^3}\frac{\chi_2 \bar{\mu}_2}{\chi_1}$},  and $\bar{V}_4=\sum_i\frac{\psi_i(2-\psi_i)}{1-\psi_i}$. Additionally, we obtain \update{$\dot{V}_3\leq -2k_1\mu_2(\sqrt{V_3})^2$, and   $\dot{V}_4\leq -\chi_1(\sqrt{\bar{V}_4})^2$.} 
 Let $\bm{x}=\begin{bmatrix}\sqrt{V_3}& \sqrt{\bar{V}_4}    \end{bmatrix}^\intercal $
Finally, we obtain\update{
\begin{align}
&\dot{\mathcal{V}}_2(\bm{x})\leq-\bm{x}^\intercal
\Lambda\bm{x},\quad \Lambda=\begin{bmatrix} 2(1-d)k_1\mu_2& \frac{dh}{2}\\ \frac{dh}{2}& \frac{d\chi_1}{\epsilon}    \end{bmatrix},
\end{align}}
where $\Lambda$ 
is a positive definite matrix for all $\epsilon<\epsilon^*$
where $\epsilon^*$ is as given by (\ref{espstarnoncomoplete}).
Hence, according to \textbf{Theorem 5.1} in \cite{Kokotovi1986SingularPM}, the origin of unforced  system $h(t,\epsilon,\bm{\theta},\bm{\psi},0)$ is uniformly asymptotically stable.
\end{proof}
\subsection{Convergence of the Swarm Source Seeking and  Level Curve Tracking under  Complete Graphs}\label{convergence of z}
In this section, we study  the convergence of the swarm  to either the source location or desired level curve. In particular,  we 
analyze the trajectory of the field measurement at the center of the swarm.
Define $z_c^d=z_c-z^d$. This implies that $z_c^d=0$ if and only if $z_c=z^d$. Taking the time derivative, $\dot{z}_c^d=\dot{z}_c=\langle\nabla z_c,\dot{\bm{r}}_c \rangle $. \update{But, using (\ref{eq:sloworiginal1}) for the complete graph case, $\dot{\bm{r}}_c=\frac{1}{M}\sum_k[k_1 (z_k-z^d)\bm{n}+k_2\bm{q}]=
k_1 (z_a-z^d)\bm{n}+k_2\bm{q}$ where $z_a=\frac{1}{M}\sum_k z_k$ is the average field measurement.
Note that , using (\ref{eq:field2}), $z_a=\frac{1}{M}\sum_k z_k=\frac{1}{M}\sum_{k=1}^M[z_c+\langle\nabla z_c ,\bm{r}_k-\bm{r}_c \rangle+\nu_k]=z_c+\nu$ where $\nu=\frac{1}{M}\sum_{k=1}^M\nu_k$ and $\frac{1}{M}\sum_{k=1}^M\langle\nabla z_c ,\bm{r}_k-\bm{r}_c \rangle=0$. That is the difference between the average and center measurements equals to the average of higher-order terms.
Then we obtain 
 \begin{align}\label{Zcd_dynamics}
&\dot{z}_c^d=k_1\|\nabla z_c\|(z_c^d
+\nu)\langle\bm{N} ,\bm{n}\rangle+k_2\|\nabla z_c\|\langle\bm{N} ,\bm{q}\rangle,
\end{align} 
where $\nu=z_a-z_c$.
Note that, even when $\langle\bm{N} ,\bm{n}\rangle=\pm1$ which implies $\langle\bm{N} ,\bm{q}\rangle=0$, $z_c^d=0$ is not an equilibrium to (\ref{Zcd_dynamics}) due to the existence of $\nu$. In the following, we present a boundedness result for the trajectory $z_c^d(t)$.
\begin{theoremEnd}{thm}\label{'lemma_z_completeLCT'}
 Suppose $\|\nabla z(\bm{r}_c)\|>\mu_3$ where $\mu_3>0$ is a constant. Furthermore, suppose $-1\leq\langle\bm{N} ,\bm{n}\rangle\leq-\epsilon_3$ and $|\nu|\leq\bar{\nu}$, where $\epsilon_3\in(0,1)$ and  $\bar{\nu}>0$ are constants. Then, the solutions of (\ref{Zcd_dynamics}) are uniformly ultimately bounded.
\end{theoremEnd}

\begin{proof}
Let $V_5:\bm{R}\to\bm{R}$ be a Lyapunov candidate function defined by $V_5=\frac{1}{2}(z_c^d)^2$, where $V_5=0$ if and only if $z_c^d=0$. 
Then we obtain
\begin{align}\label{V5dot}
    &\dot{V}_5\leq -2k_1\mu_3\epsilon_3(1-\epsilon_3) V_5,\forall|z_c^d|\geq\frac{k_1\bar{\nu} +k_2\sqrt{1-\epsilon_3^2}}{k_1\epsilon_3^2 },
\end{align}
where $\dot{V}_5=0$ if and only if $z_c^d=0$. Let $\alpha_5(|z_c^d|)=\alpha_6(|z_c^d|)=\frac{1}{2}|z_c^d|^2$ be class $\mathcal{K}$ functions. Then $\alpha_5(|z_c^d|)\leq V_5\leq\alpha_6(|z_c^d|)$.
Therefore,  according to \textbf{Theorem 4.18} in \cite{khalil2002nonlinear}, the  trajectories of the system (\ref{Zcd_dynamics}) are uniformly ultimately bounded.
\end{proof} 
Note that, (\ref{V5dot}) implies that the $z_c^d$ trajectories of (\ref{Zcd_dynamics}) will converge to a strip around the desired level curve and the strip is defined by $\{\bm{r}_c||z_c^d|\leq\frac{k_1\bar{\nu} +k_2\sqrt{1-\epsilon_3^2}}{k_1\epsilon_3^2 }\}$. If we only consider source seeking, i.e. $k_2=0$ and $z_c^d=z_c$, then the $z_c$ 
trajectories of (\ref{Zcd_dynamics}) will converge to a neighborhood around the source location defined by $\{\bm{r}_c||z_c^d|\leq\frac{\bar{\nu}}{\epsilon_3^2}\}$.
\begin{remark} 
Theorem~\ref{'lemma_z_completeLCT'} can be viewed as a general result. In particular, if each agent moves according to  $\dot{\bm{r}}_i=k_1(z(\bm{r}_i)-z^d)\bm{v}+k_2\bm{v}^\perp$,  then as long as $-1\leq\langle\bm{N} ,\bm{v}\rangle\leq-\epsilon_3$,  the center of the swarm will converge to a neighborhood of the source location or to a strip along the desired level curve. The size of these neighborhoods is determined by $k_1$, $k_2$, and $\bar{\nu}$. The direction $\bm{v}$ could be along the exact negative gradient, or along a sub-gradient direction, i.e. $\bm{v}$ satisfies $z(\bm{r}_i)-z(\bm{r}_c)\geq\langle\bm{v},\bm{r}_i-\bm{r}_c\rangle$. In this paper,  $\bm{v}$ is given by the PCA direction $\bm{n}$  which is proved  in \textbf{Corollary}~\ref{'corollary1'} to satisfy the required assumption  $-1\leq\langle\bm{N} ,\bm{n}\rangle\leq-\epsilon_3<0$.
\end{remark}
}


\vspace{-15pt}
\subsection{Convergence of the Swarm  for Source Seeking under Incomplete Graphs}\label{convergence of zincomplete}

Using (\ref{eq:sloworiginal1}) for $z^d=0$ and $k_2=0$, we obtain
\begin{align}\label{rcdotincomplete}
    &\dot{\bm{r}}_{c,i}=k_1z_{a,i}\bm{n}_i+\frac{k_1}{M_i}\sum_{j\in\mathcal{N}_i}z_j(\bm{n}_j-\bm{n}_i),
\end{align}
where $z_{a,i}=(1/M_i)\sum_{k\in\mathcal{H}_i}z_i$, $\bm{r}_{c,i}=(1/M_i)\sum_{k\in\mathcal{H}_i}\bm{r}_k$, and $M_i=|\mathcal{H}_i|=1+|\mathcal{N}_i|$.
\update{
Let $\nu_{c,i}=\frac{1}{M_i}\sum_{k\in\mathcal{H}_i}\nu_k$. Hence, using (\ref{eq:field2})  $\nu_{c,i}=z_{a,i}-z_{c,i}$, i.e. the difference between the local average and center measurements of the field. Let
\begin{align}\label{ei}
&\bm{e}_i=\frac{1}{M_i}\sum_{j\in\mathcal{N}_i}z_j(\bm{n}_j- \bm{n}_i).    
\end{align}
Then, substituting (\ref{ei}) into  (\ref{rcdotincomplete}), we obtain
\begin{align}\label{Zci_dynamics}
&\dot{z}_{c,i}=k_1\|\nabla z_{c,i}\|(z_{c,i}+\nu_{c,i})\langle\bm{N}_{c,i} , \bm{n}_i\rangle+k_1\|\nabla z_{c,i}\|\langle\bm{N}_{c,i} ,\bm{e}_i\rangle,
\end{align} 
Define  $\bm{z}_c=[z_{c,1},\cdots,z_{c,M}]^\intercal$. Then let $\dot{\bm{z}}_c=[\dot{z}_{c,1},\cdots,\dot{z}_{c,M}]^\intercal$, where for each agent $\dot{z}_{c,i}$ is given by (\ref{Zci_dynamics}). In the following, we present a boundedness result for the trajectory $\bm{z}_c(t)$.
\begin{theoremEnd}{proposition}
Suppose for each agent, $\|\nabla z_{c,i}\|>\mu_4$,  $-1\leq\langle\bm{N}_{c,i} , \bm{n}_i\rangle\leq-\epsilon_4$, $|\nu_{c,i}|\leq\bar{\nu}$,  and $\langle\bm{N}_{c,i} , \bm{e}_i\rangle\leq\bar{e}$, where  $\mu_4>0$, $\epsilon_4>0$, $\bar{\nu}>0$ and $\bar{e}>0$ are constants. Then, the solutions of (\ref{Zci_dynamics}) are uniformly ultimately bounded.
\end{theoremEnd}
\begin{proof}
Let $V_6:\bm{R}\to\bm{R}$ be a Lyapunov candidate function defined by $V_6=\frac{1}{2}\sum_{i=1}^Mz_{c,i}^2=\frac{1}{2}\|\bm{z}_c\|^2$, where $V_6=0$ if and only if $\bm{z}_c=\bm{0}$, i.e. $z_{c,i}=0$ $\forall i$. 
Then we obtain
\begin{align}\label{v6dott}
    &\dot{V}_6
    \leq -2k_1\mu_4\epsilon_4(1-\epsilon_4)V_6,\quad\forall \|\bm{z_c}\|\geq\frac{(\bar{\nu}+\bar{e})}{\epsilon}\sqrt{M}
\end{align}
where $\dot{V}_6=0$ if and only if $\bm{z}_c=\bm{0}$, and $M$ is the total number of agents.
Let $\alpha_7(\|\bm{z}_c\|)=\alpha_8(\|\bm{z}_c\|)=\frac{1}{2}\|\bm{z}_c\|^2$ which are class $\mathcal{K}$ that satisfy: $\alpha_7(\|\bm{z}_c\|)\leq V_6\leq\alpha_8(\|\bm{z}_c\|)$.
Therefore,  according to \textbf{Theorem 4.18} in \cite{khalil2002nonlinear}, the trajectories of $\dot{\bm{z}}_c$ are uniformly ultimately bounded.
\end{proof} 
Note that the assumption that $-1\leq\langle\bm{N}_{c,i} , \bm{n}_i\rangle\leq-\epsilon_4<0$ is satisfied by 
\textbf{Corollary}~\ref{'corollary1_noncomplete'}. 
This result implies that, as seen from (\ref{v6dott}), the trajectories of $\bm{z}_c$, and hence the agents, converge to a neighborhood defined by $\{\bm{r}_c|\|\bm{z}(\bm{r}_c)\|\geq\frac{(\bar{\nu}+\bar{e})}{\epsilon}\sqrt{M}\}$. The bound $\bar{e}$, as seen from (\ref{ei}), is determined by the connectivity of the graph. A more connected graph increases the chance that the local PCA directions $\bm{n}_i$ being aligned with each other. 
}


\vspace{-13pt}
 \section{Simulation and Experimental Results}\label{Results}
 In this section, we validate the proposed model through computer simulation and  physical experiments. We used two robotic platforms: the  Georgia Tech Robotarium  mobile robots \cite{pickem2017robotarium}, and the flying Georgia Tech Miniature Autonomous Blimps \cite{cho2017autopilot}. In what follows, we first present the source seeking results and then the level curve tracking results. In all simulations and experiment, we set $\epsilon=0.01$ in   (\ref{eq:fastoriginal}). This means that we run the PCA flow (\ref{eq:PCAflow}) for a time $\tau=\frac{dt}{\epsilon}$ where $dt=0.01$ is the step time used to update (\ref{eq:controllaw}). Additionally, for convex fields, we use the field function $z(\bm{r}_i)=\|\bm{r}_i\|^2$. For non-convex fields, we use the field function  $z(\bm{r}_i)=2-\exp(-(\bm{r}_i-\bm{a})^\intercal S_1(\bm{r}_i-\bm{a}))-\exp(-(\bm{r}_i-\bm{b})^\intercal A^\intercal S_2A(\bm{r}_i-\bm{b}))+\|\bm{r}_i\|$, where $\bm{a}=[1,0]^\intercal$, $\bm{b}=[0,-2]^\intercal$, $S_1=0.9\begin{bmatrix}(1/\sqrt{30}) &0\\0 & 1 \end{bmatrix}$, $S_2=0.9\begin{bmatrix}1 &0\\0 & (1/\sqrt{15}) \end{bmatrix}$ and $A=(\sqrt{2}/2)\begin{bmatrix}1 &-1\\1 & 1 \end{bmatrix}$.
 \vspace{-13pt}
\subsection{Source Seeking}
\subsubsection{Simulation Results}
We simulated \textbf{Algorithm 1} in virtual scalar convex and non-convex $2$-D fields for swarms that have complete and incomplete  connectivity graphs.  In all the simulations, we set $k_1=1$  in (\ref{eq:controllaw}). Additionally, the source is located at the origin. In all of the following figures, bold blue discs represent the agents and the blue arrows indicate the direction $\bm{n}_i$, where they turned to red color at the end of the simulation. The lines connecting the agents represent the edges of the network and the pink paths represent the trajectories of the agents. The contour lines represent the level curves of the field.
\ifdefined\PICTURES
\begin{figure}[h!]
     \centering
      \includegraphics[scale=0.4]{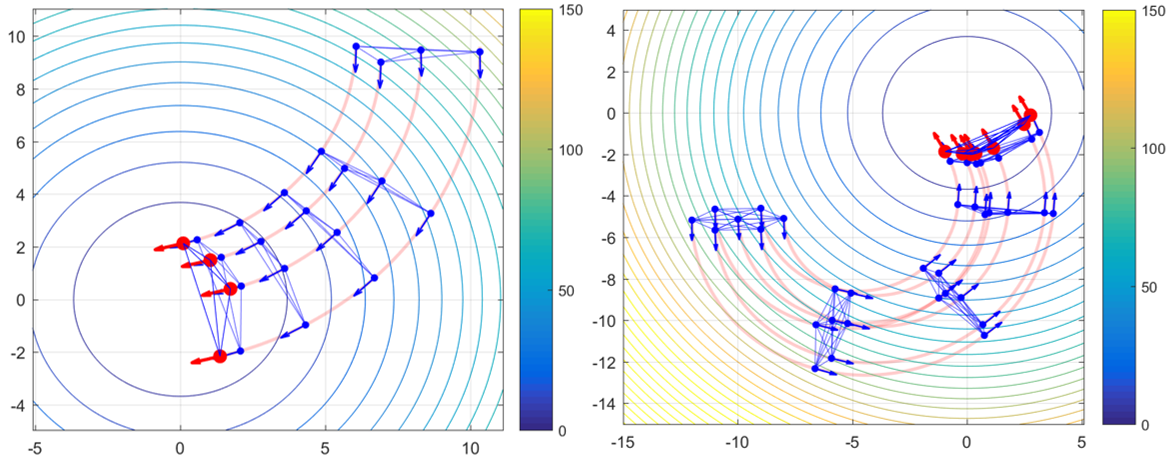}
      \caption{Swarms of $4$ agents (left) and $7$ agents (right) in  complete graphs.}
      \label{Sim1}
\end{figure}\vspace{-13pt}
\fi
In Fig.~\ref{Sim1},  swarms of $4$ agents (left) and $7$ agents (right) in a complete graph are used to locate a convex field starting from different initial positions.  
As predicted by \textbf{Theorem~\ref{'coupledsystemcomplete'}}, the two swarms successfully steered towards the source even though the $7$-agent swarm was initially heading towards the positive direction of the field gradient. Additionally, as predicted by \textbf{Lemma~\ref{'lambdaqdot is zero'}}, the variance $\lambda^q$ is constant while $\lambda^n$ is varying.
\ifdefined\PICTURES
\begin{figure}[h!]
     \centering
      \includegraphics[scale=0.41]{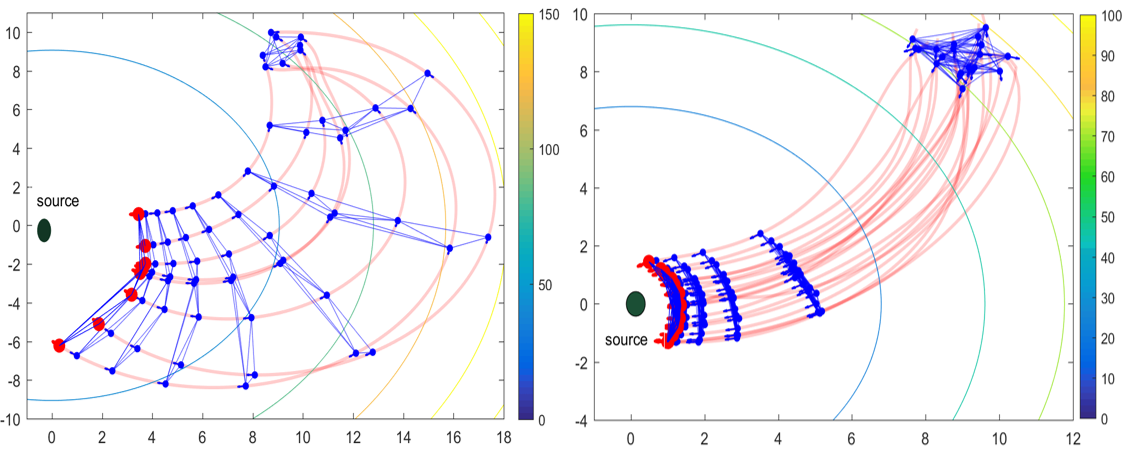}
      \caption{Swarms of $8$ agents (left) and $20$ agents (right) in incomplete graphs}
      \label{Sim2}
\end{figure}
\fi
\\
In the $8$-agent swarm in the left of Fig.~\ref{Sim2}, the connectivity graph is incomplete. Hence each agent applies the PCA flow locally resulting in different $\bm{n}_i$, as it clear by the blue arrows at the initial time. Nevertheless, as predicted by  \textbf{Proposition~\ref{'coupledsystemnoncomplete'}}, the swarm steers towards the source and each $\bm{n}_i$ converges to the negative direction of the local gradient $\bm{N}_{c,i}$. However, since initially for some agents $\langle\bm{n}_i,\bm{N}_{c,i}\rangle>0$, the swarms disperse significantly. To save the connectivity of the graph when it is incomplete, in the $20$-agent swarm in the right of Fig.~\ref{Sim2}, modify the control law (\ref{eq:controllaw}) as 
\begin{align}\label{eq:controllawmodified}
&\bm{u}_i(t)=k_1(z_i(t)-z^d)\bm{n}_i+k_2\bm{q}_i+k_f\bm{v}_{i,\bm{q}},
\end{align}
where $k_f$ is a constant gain, and $v_{i,\bm{q}}$ is designed as follows  
\begin{align}\label{eq:formation}
&\bm{v}_{i,\bm{q}}=\sum_{j\in\mathcal{N}_i}(\langle \bm{r}_j-  \bm{r}_i,\bm{q}_i\rangle-d_{ij})\langle \bm{r}_j-  \bm{r}_i,\bm{q}_i\rangle\bm{q}_i,
\end{align}
which is to maintain a desired distance $d_{ij}$ only along the $\bm{q}_i$ direction \cite{al2018integrating}. As shown in the right of Fig.~\ref{Sim2}, each $\bm{n}_i$ converges to  $-\bm{N}_{c,i}$, but also, the agents in the swarm keep close to each other due to (\ref{eq:formation}).
\ifdefined\PICTURES
\begin{figure}[h!]
     \centering
      \includegraphics[scale=0.37]{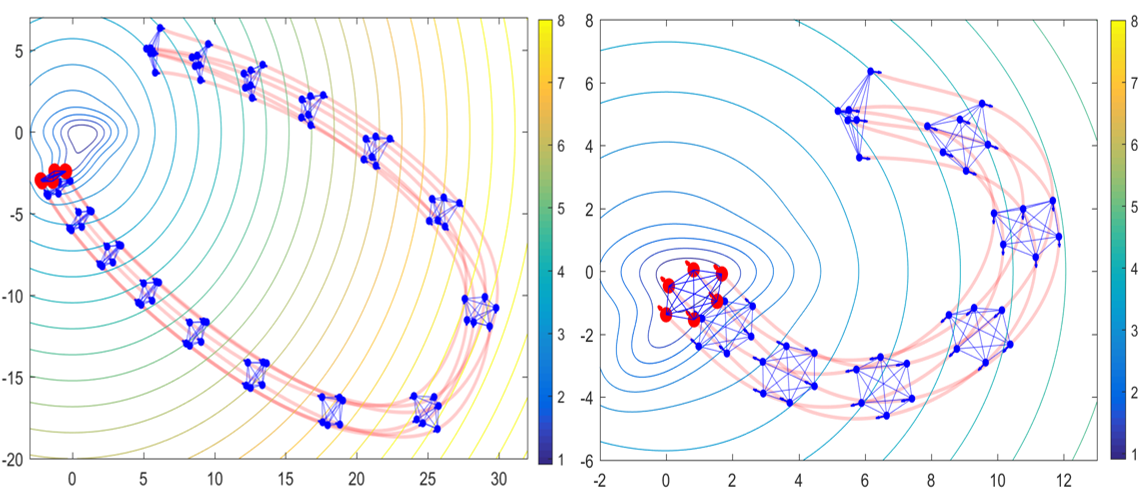}
      \caption{Complete network of $6$ agents in a non convex field}
      \label{Sim3}
\end{figure}
\fi
\\
As suggested by (\ref{eq:ndot6complete22}), $\bm{n}$ changes faster when $(\lambda^q-\lambda^n)$ is small. To justify this,  in Fig.~\ref{Sim3}, a swarm of $6$ agents and a complete graph is simulated in a non-convex field. In the left, we added (\ref{eq:controllawmodified}) to maintain a distance only along $\bm{q}$. However, in the right we maintain a  distance  along both directions by adding $\bm{v}_{i,\bm{n}}$ to (\ref{eq:controllawmodified}), where $\bm{v}_{i,\bm{n}}$ is obtained by replacing $\bm{q}$ by $\bm{n}$ in (\ref{eq:formation}). 
Although the two swarms start at the same location, since the one in the right maintained smaller $(\lambda^q-\lambda^n)$,  it steered faster towards the source than the one in the left which took a long distance to turn.  This intuitively reveals the effect of the different formation schemes. In particular, a swarm with a larger spatial distribution encodes more diverse information about the field and hence the swarm steers faster towards the source.
\ifdefined\PICTURES
\begin{figure}[h!]
     \centering
      \includegraphics[scale=0.37]{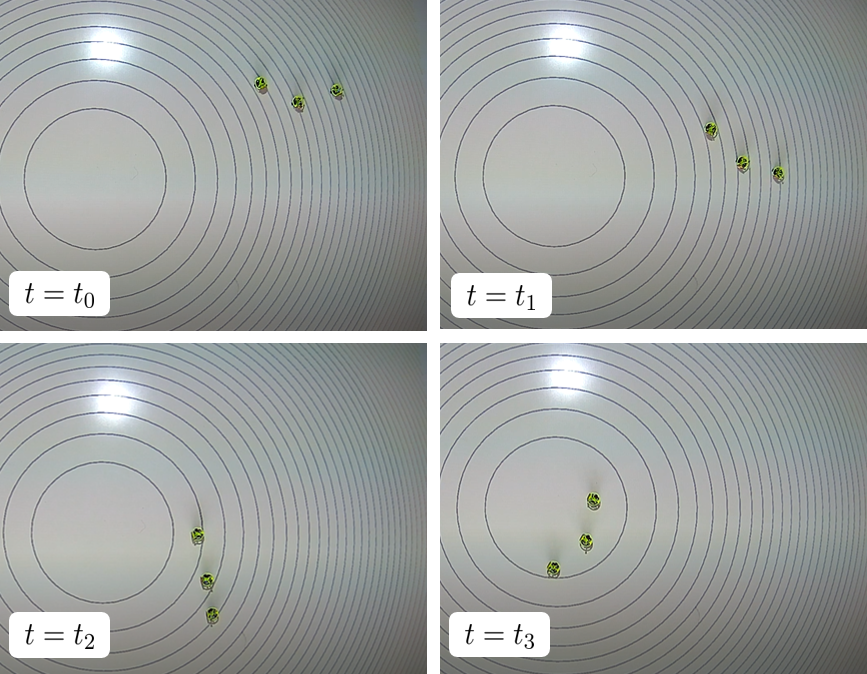}
      \caption{A $3$-agent swarm in an incomplete graph}
      \label{Robo1}
\end{figure}

\begin{figure}[h!]
     \centering
      \includegraphics[scale=0.27]{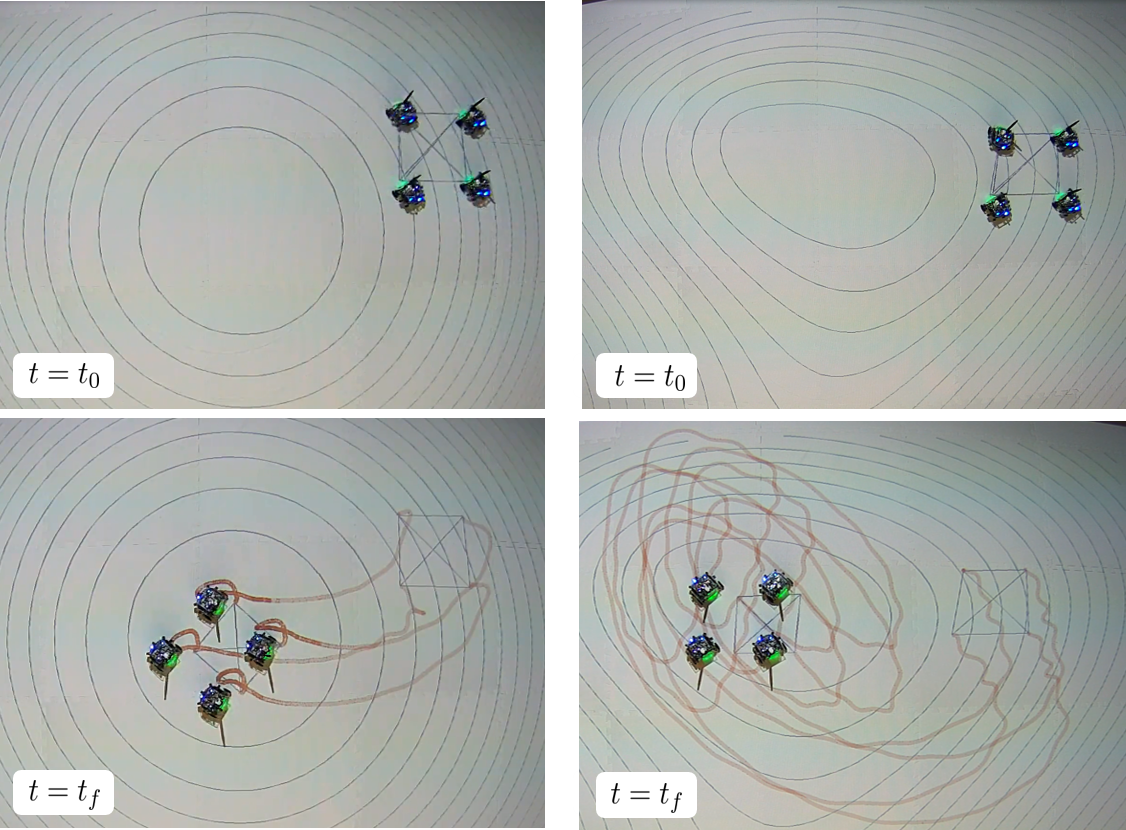}
      \caption{A $4$-agent swarm in  convex (left) and non-convex fields (right)}
      \label{Robo2}
\end{figure}\vspace{-10pt}
\fi
\vspace{-15pt}
\subsubsection{Experimental Results}
We tested \textbf{Algorithm 1} using unicycle mobile robots at the Georgia Teach Robotarium \cite{pickem2017robotarium}. The diameter of each robot is about $0.05m$ and the dimensions of the experimental space are about $2m\times3m$.  Since these robots do not have light sensors, we used virtual fields where we projected their level curves on the surface for a visualization purpose. In Fig.~\ref{Robo1}, we show the results of a $3$-robot system in a convex field. The graph is complete, however, without formation. Alternatively, we show in Fig.~\ref{Robo2} the results of a $4$-robot system in a complete graph and with formation. The two swarms start at the same location, however, since the field is convex in the left, the swarm approaches the source and quickly settle. However, since the field in non-convex in the right, the swarm requires more time to eventually settle at the source.

Alternatively, we installed light sensors in the Georgia Tech Miniature Blimps \cite{cho2017autopilot} and then performed source seeking in a physical light field.  The diameter of each blimp is about $0.7$ m and the dimensions of the experimental space are about $4$ m $\times$ $4$ m. To make the minimum at the source, we inverted the field by using $\frac{1}{z_i}$ instead of $z_i$. Snapshots of two experiments are shown in Fig.~\ref{EXPBlimp3} and  Fig.~\ref{EXPBlimp4}, where initially  in the former $\langle\bm{N},\bm{n}\rangle<0$, and in the latter $\langle\bm{N},\bm{n}\rangle>0$.  
\ifdefined\PICTURES
\begin{figure}[h!]
     \centering
      \includegraphics[scale=0.52]{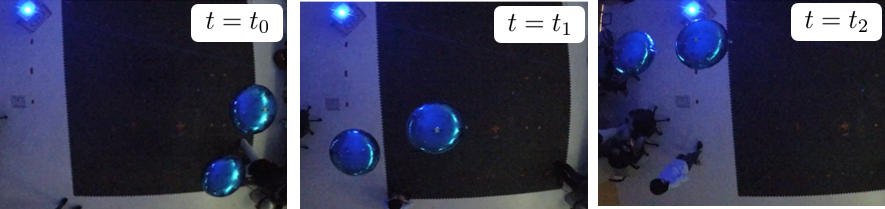}
      \caption{The two blimps  initially have $\langle\bm{N},\bm{n}\rangle<0$. }
      \label{EXPBlimp3}
\end{figure}
\begin{figure}[h!]
     \centering
      \includegraphics[scale=0.52]{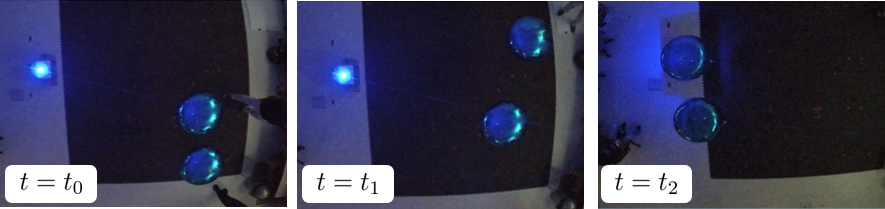}
      \caption{The two blimps  initially have $\langle\bm{N},\bm{n}\rangle>0$.}
      \label{EXPBlimp4}
\end{figure}
\fi
Additionally, we presented in   Fig.~\ref{EXPBlimp1} the trajectories of these two experiments, where the one on the lest corresponds to Fig.~\ref{EXPBlimp3}, and the other one corresponds to Fig.~\ref{EXPBlimp4}. The trajectories are colored based on the light intensity where the black diamond is the source location and the red discs are the starting locations. Despite many messing measurements, the blimps are able to locate the source.
\ifdefined\PICTURES
\begin{figure}[h!]
     \centering
      \includegraphics[scale=0.34]{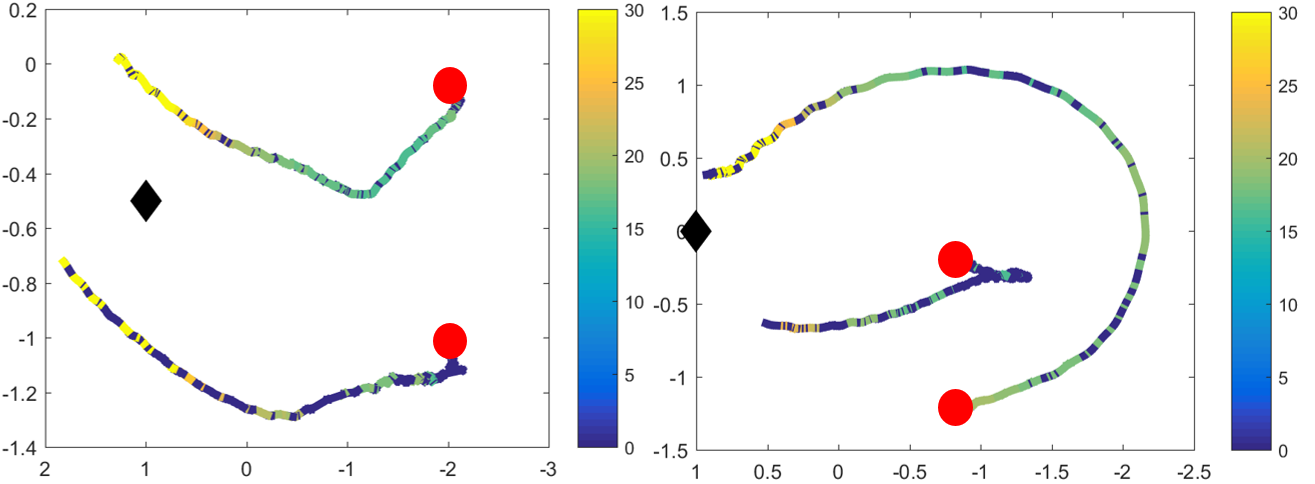}
      \caption{Trajectories of the two experiments of Fig.~\ref{EXPBlimp3} (left) and Fig.~\ref{EXPBlimp4} (right). The trajectories are colored based on the light intensity. The red discs represent the starting points and the diamond is the light source location.}
      \label{EXPBlimp1}
\end{figure}\vspace{-10pt}
\fi
\vspace{-15pt}
\subsection{Level Curve Tracking}
In what follows, we set $z^d=2$ and $\epsilon=0.01$.  Additionally, we set $k_1=2$, $k_2=0.5$.
\subsubsection{Simulation Results}
In Fig.~\ref{SimLCT7agents}, a $7$-agent system with a line graph is simulated in a convex field (left) and a non-convex field (right). Alternatively, a $10$-agent system is simulated in a convex field Fig.~\ref{SimLCTrandom},  where in the left the graph is incomplete and static, while in the right the graph is dynamic. In the dynamic graph, each agent chooses the closest three agents as its neighbors at each instant of time. Despite the graphs are arbitrary, the system is able to track the desired level curve.
\ifdefined\PICTURES
\begin{figure}[h!]
     \centering
      \includegraphics[scale=0.27]{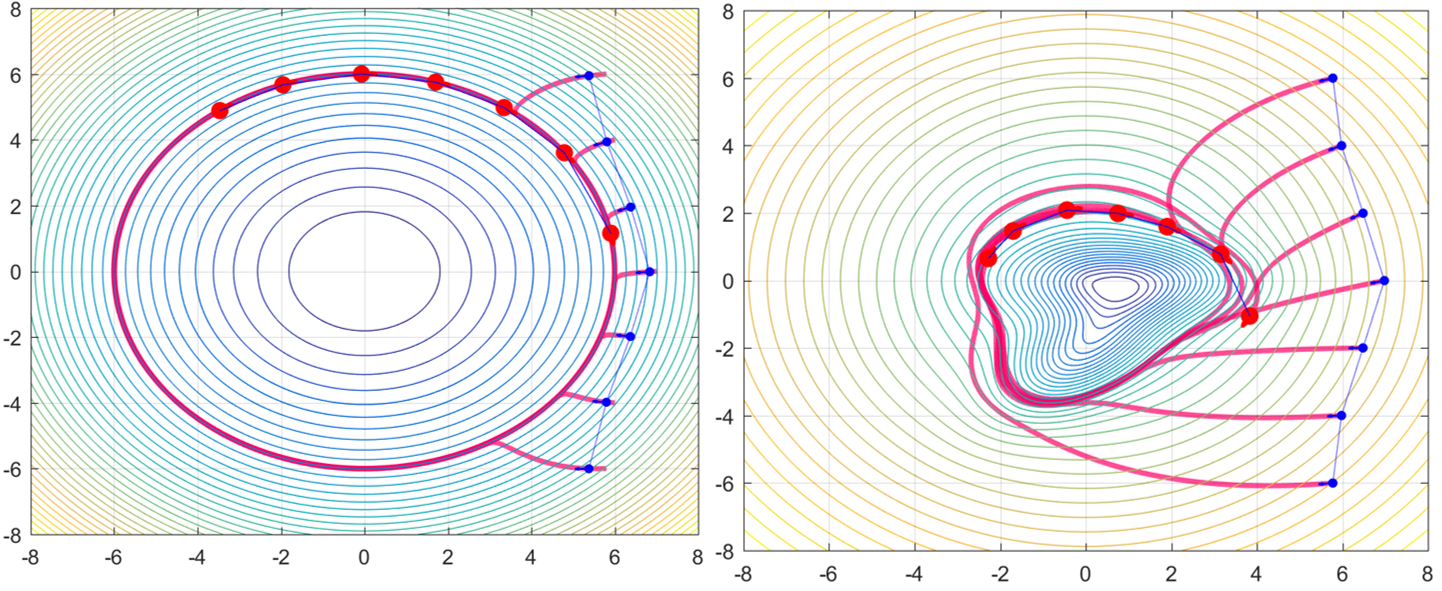}
      \caption{a $7$-agent system with a line graph in a convex field (left) and a non-convex field (right).}
      \label{SimLCT7agents}
\end{figure}
\begin{figure}[h!]
     \centering
      \includegraphics[scale=0.37]{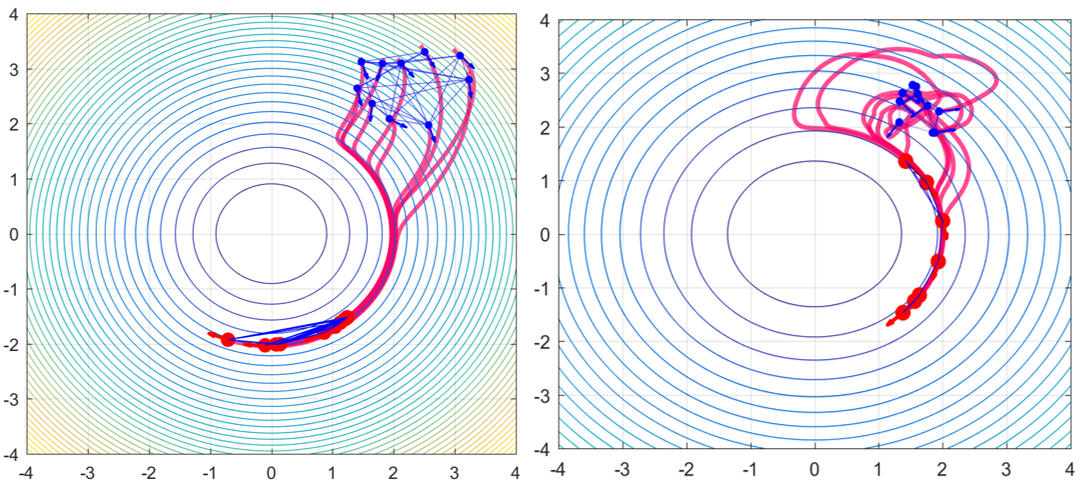}
      \caption{a $10$-agent system with a static incomplete graph  (left) and a dynamic graph (right).}
      \label{SimLCTrandom}
\end{figure}
\fi
\vspace{-13pt}
\subsubsection{Experimental  Results}
In Fig.~\ref{Robo4}, we implemented the level curve tracking algorithm using four Robotarium robots in a virtual non-convex field. The robots are connected by a line graph. Despite the lack of formation control and the simplicity of the control law, the swarm is able to track the desired level curve smoothly.
\ifdefined\PICTURES
\begin{figure}[!]
     \centering
      \includegraphics[scale=0.37]{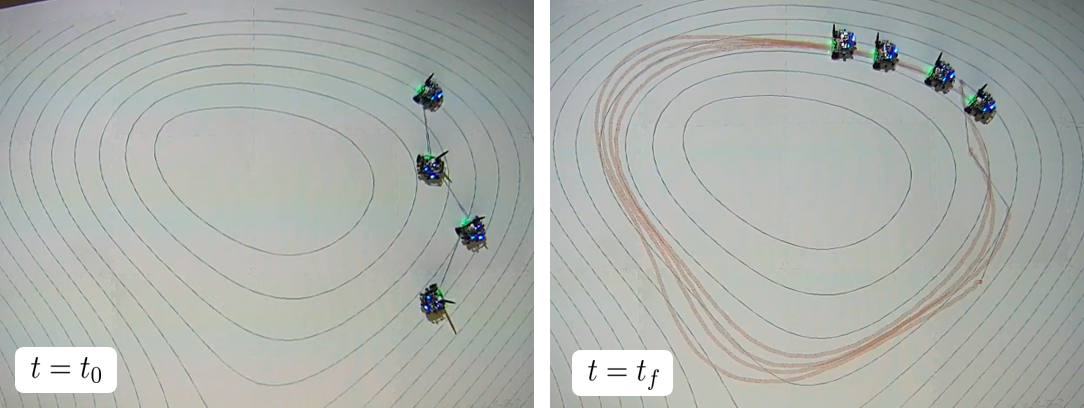}
      \caption{Four robots in a non convex field.}
      \label{Robo4}\vspace{-13pt}
\end{figure}
\fi
\section{Conclusion}\label{Conclusion}
In this paper, we proposed a distributed active perception strategy for source seeking and level curve tracking. Using the body frame obtained by PCA perception, we designed a distributed control law that enables swarms of various sizes and graph structures to perform collective source seeking and level curve tracking of scalar fields without the need to explicitly estimate the field gradient or explicitly share measurements among the agents. We obtained several stability results in a singular perturbation framework justifying the robustness and convergence of the algorithms. The simulation and experimental results suggest the efficiency and generality of the proposed model. In the future, we will design control laws for different swarm applications within the proposed framework of the distributed active perception. 
\section*{acknowledgements}
S. Al-Abri and F. Zhang were supported by ONR grants  N00014-19-1-2556 and N00014-19-1-2266; NSF
grants  OCE-1559475,  CNS-1828678, and S\&AS-1849228; NRL grants N00173-17-1-G001 and N00173-19-P-1412 ; and NOAA grant NA16NOS0120028. 
\section{proofs of the information dynamics}\label{proofssection}

\printProofs


\bibliographystyle{plain}        
\bibliography{main}
\begin{IEEEbiography}[{\includegraphics[width=1in,height=1in,clip,keepaspectratio]{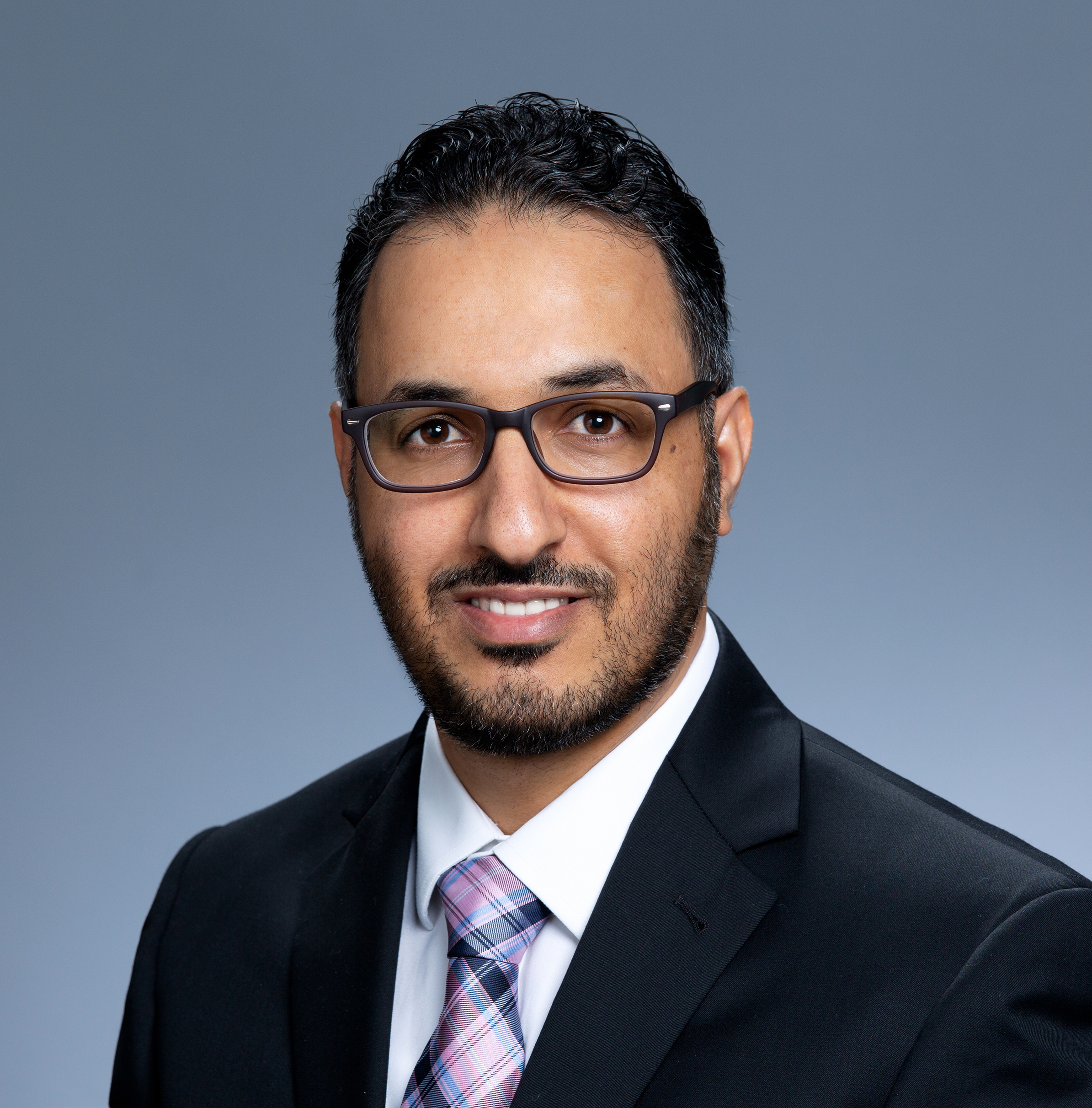}}]{Said Al-Abri}
is currently a Postdoctoral Scholar  at the School of Electrical and Computer Engineering at Georgia Institute of Technology working under the supervision of Professor Fumin Zhang. He earned his Ph.D. degree from the Georgia Institute of Technology on August 2019, M.S. degree from the University of Central Florida on May 2013, and B.S. degree from the Sultan Qaboos University on June 2009, all in Electrical and Computer Engineering. He is currently conducting research on bio-inspired distributed autonomy and optimization algorithms.
He received the Outstanding ECE Graduate Teaching Assistant Award at Georgia Tech in 2019, the Best Three-Minute Presentation at American Control Conference 2019, and the Best Poster Award at the Coordinated Science Laboratory Student Symposium, 2019.
\end{IEEEbiography}
 \begin{IEEEbiography}[{\includegraphics[width=1in,height=1in,clip,keepaspectratio]{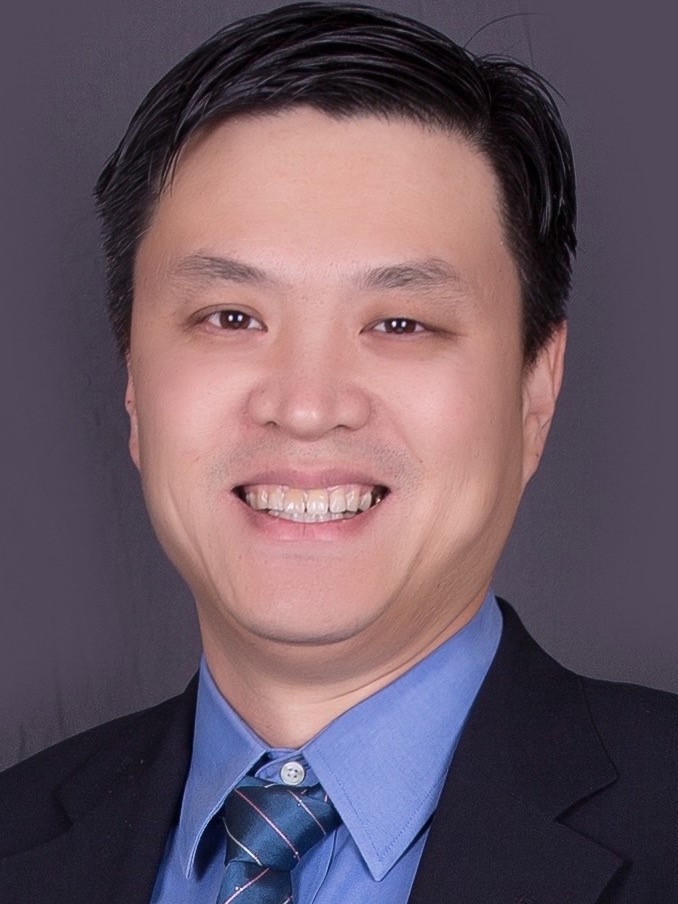}}]{Fumin Zhang}
received the B.S. and M.S. degrees from Tsinghua University, Beijing, China, in 1995 and 1998, respectively, and the Ph.D. degree from the Department of Electrical and Computer Engineering, University of Maryland, College Park, in 2004. He joined the School of ECE, Georgia Institute of Technology in 2007, where he is a Professor. He was a Lecturer and Postdoctoral Research Associate in the Mechanical and Aerospace Engineering Department, Princeton University from 2004 to 2007. His research interests include marine autonomy, mobile sensor networks, and theoretical foundations for battery supported cyber-physical systems. He received the NSF CAREER Award in 2009, and the ONR YIP Award in 2010.
\end{IEEEbiography}



\end{document}